\documentclass[fleqn,usenatbib]{mnras}

\usepackage[T1]{fontenc}

\usepackage{newtxtext,newtxmath}
\usepackage{subfigure}

\usepackage{graphicx}	% Including figure files
\usepackage{amsmath}	% Advanced maths commands

\usepackage{float}
\usepackage{arydshln}
\title[A radiative transfer model for W2246$-$0526]
{\textit{JWST} observations and a model for the extremely luminous obscured quasar W2246$-$0526 at \boldmath$z=4.6$}

\author[Charalambia Varnava] {\parbox{\linewidth}{Charalambia Varnava$^{1,2}$\thanks{E-mail: \href{mailto:varnava.haris@gmail.com}{varnava.haris@gmail.com} (CV)}, 
Andreas Efstathiou$^{1}$, Tanio D{\'\i}az-Santos$^{3,1}$ and Duncan Farrah$^{4,5}$
}
\\
\\
$^{1}$School of Sciences, European University Cyprus, Diogenes street, Engomi, 1516 Nicosia, Cyprus\\
$^{2}$CaSToRC, The Cyprus Institute, Konstantinou Kavafi street, Aglantzia, 2121 Nicosia, Cyprus\\
$^{3}$Institute of Astrophysics, Foundation for Research and Technology-Hellas (FORTH), Heraklion, 70013, Greece\\
$^{4}$Department of Physics and Astronomy, University of Hawaii, 2505 Correa Road, Honolulu, HI 96822, USA\\
$^{5}$Institute for Astronomy, University of Hawaii, 2680 Woodlawn Drive, Honolulu, HI 96822, USA
}

\date{Accepted 2026 XX; Received 2026 YY; in original form 2025 ZZ}

\pubyear{2026}

\begin{document}
\label{firstpage}
\pagerange{\pageref{firstpage}--\pageref{lastpage}} 
\maketitle

\begin{abstract}
We present new \textit{JWST}/MIRI-MRS data of the $z=4.601$ extremely luminous obscured quasar WISEA J224607.56$-$052634.9 (W2246$-$0526). Our fits of its spectral energy distribution (SED) with the SED fitting code SMART (Spectral energy distributions Markov chain Analysis with Radiative Transfer models) predict an active galactic nucleus (AGN) fraction in the range $72-81$ per cent, an intrinsic AGN luminosity of $4.2-7.2 \times 10^{14} L_\odot$, a polar dust luminosity of $1.6-1.7 \times 10^{14} L_\odot$, a black hole mass of $1.3-2.3 \times 10^{10} M_\odot$ (assuming the quasar is accreting at the Eddington limit), a star formation rate (SFR) of $360-2900~M_\odot yr^{-1}$ and a stellar mass of $4.8-5 \times 10^{11} M_\odot$. The stellar and black hole masses of W2246$-$0526 are typical of a giant elliptical galaxy at $z=0$.
We find statistically significant evidence for the presence of a hot dust component, which we interpret as polar dust in the context of a torus geometry, based on recent results obtained for nearby AGN. We explore two smooth and two two-phase models for the AGN torus, to put constraints on the AGN fraction of the galaxy, the black hole mass and its SFR. We show that the presence of polar dust affects the estimate of the AGN luminosity and we recommend to take into account this component in SED fits of other high-redshift obscured AGN/quasars. Despite the large difference in luminosity, we discuss possible links between the presence of this hot dust component in W2246$-$0526 and in some local AGN, suggesting that they may have a different origin.
\end{abstract}


\section{Introduction}\label{sec:intro}

Recent mid-infrared surveys have revealed a significant population of high-redshift sources, whose spectral energy distributions (SEDs) peak in the mid-infrared. These sources have been named hot dust-obscured galaxies (Hot DOGs; \citealt{wright10,eins12,wu12,bridge13,tsai15,far17}). Hot DOGs are hyperluminous infrared galaxies (HLIRGs), with $1-1000~\mu m$ luminosities that exceed $10^{13} L_\odot$, identified by their distinct mid-infrared colors in Wide-field Infrared Survey Explorer (\textit{WISE}; \citealt{wright10}) data. They can be extremely luminous, with infrared luminosities exceeding $10^{14} L_\odot$, raising important questions. How can these systems reach such extreme luminosities and what is the origin of their emission?

It is widely believed that Hot DOGs are mainly powered by an active galactic nucleus (AGN) with evidence of super-Eddington accretion (e.g. \citealt{tsai18}). They therefore represent important laboratories for the study of the AGN obscurer model at high redshifts, as the dilution from starlight and star formation is minimized. These systems also offer the possibility of probing the properties of polar dust, which appears to be present in a significant fraction of low-redshift AGN (e.g. \citealt{asm19}), but is usually much fainter than the torus itself and difficult to study either with high-resolution observations or SED fitting \citep{efs22,smart,varn25}.

The presence of polar dust may be related to feedback processes going on in galaxies at all redshifts, which are believed to play a significant role in regulating and/or quenching star formation in galaxies. For example, polar dust has been suggested to indicate dusty winds powered by AGN radiation pressure \citep{hk17,honig19}. Alternatively, polar dust may be related to molecular clouds and/or shocked regions in the narrow-line region \citep{lopez25}. \cite{lyu22} presented correlations between the $\lambda>5 \mu m$ emission of AGN and the forbidden line emission, suggesting that a significant fraction of the mid-infrared emission comes from the narrow-line region. Although the nature of polar dust may still be unclear, SED fitting with this component may be a powerful technique to reveal its presence at any redshift. 

The prototypical Seyfert 2 galaxy NGC~1068 was the first AGN to be recognized by its mid-infrared emission to have polar dust. \cite{braatz93} and \cite{cam93} detected extended mid-infrared emission, which was found to align with the narrow-line region of the AGN. \cite{efs95} incorporated a polar dust component with a density distribution following an inverse square law with radius in their model for the nucleus of NGC~1068. The existence of polar dust in AGN is now supported by high-resolution mid-infrared imaging of local Seyfert galaxies (e.g. \citealt{jaffe04,raban09,tristram09,tristram14,honig12,honig13,burt13,lopez14,lopez16b,lopez16a,stal17,stal19,leftley18,asm19,gamez22,isbell22,isbell23,lopez25}).

Recent results from SED fitting of a number of luminous infrared galaxies (LIRGs) and ultraluminous infrared galaxies (ULIRGs), with $1-1000~\mu m$ luminosities that exceed $10^{11}$ and $10^{12}  L_\odot$ respectively, also support the idea of the presence of dust in the polar regions of at least some AGN \citep{matt18,efs22,reyn22, smart,varn25}. \cite{smart} and \cite{varn25} fitted with SMART\footnote{\url{https://github.com/ch-var/SMART}} (Spectral energy distributions Markov chain Analysis with Radiative Transfer models; \citealt{smart,ascl}) the HERschel Ultraluminous Infrared Galaxy Survey (HERUS) sample \citep{far13} of 42 local ULIRGs and found evidence for polar dust in IRAS~05189-2524, IRAS~07598+6508 and IRAS~13451+1232 (about 7 per cent of the sample), as the addition of the polar dust model significantly improved the fits. The galaxy IRAS~05189-2524 was recognized from the mid-90s as a ULIRG that has characteristics similar to those of the prototypical Seyfert 2 galaxy NGC~1068. In particular, it was found by \cite{young96} to show broad lines in polarized flux. In all of these recent SED models, polar dust is assumed to be hot (at a constant temperature of $\sim$ 940$-$1050 K) and concentrated in optically thick clouds.

WISEA J224607.56$-$052634.9 (W2246$-$0526) at $z=4.601$ is the most distant and luminous Hot DOG discovered so far, with a bolometric luminosity exceeding $3 \times 10^{14} L_\odot$, placing it well into the extremely luminous infrared galaxy (ELIRG) range \citep{tsai15}. According to \cite{tsai15}, the SED of W2246$-$0526 is dominated by hot dust with a temperature higher than 450K, indicative of a dominant AGN and an accreting supermassive black hole (SMBH). \cite{tsai18} re-evaluated the bolometric luminosity of W2246$-$0526, considering the possible contribution of a nearby foreground galaxy. The updated estimate, based on power-law interpolation of the well-sampled SED, is $3.6 \times 10^{14} L_\odot$. These estimates of the AGN luminosity do not take into account the anisotropy of the emission of the AGN torus, which requires a correction to the observed luminosity of the AGN to determine the intrinsic luminosity \citep{efstathiou06,efs22,smart,varn24,varn25}. In this paper, we present new \textit{JWST} data for W2246$-$0526 and explore with SED fitting models the origin of the emission of this system.

Various AGN torus models and SED fitting codes are documented in the literature. However, it remains an open question how well we can estimate the star formation rate (SFR) and the AGN luminosity of galaxies that host obscured quasars at $z>4$. This paper aims to quantify these uncertainties, employing our recently developed Bayesian SED fitting code SMART. \cite{far26} recently used SMART to explore how accurately obscured galaxy luminosities can be determined from near- to far-infrared SED fitting. SMART can fit an SED with eight different combinations of radiative transfer models, each exploring a different AGN torus model \citep{efstathiou95,fritz06,sieb15,stal16}, along with a starburst \citep{efstathiou00,efstathiou09} and a host galaxy model \citep{efs21}, which can be spheroidal or disc. The method also allows the inclusion of a polar dust component associated with the AGN \citep{efstathiou06}. This approach allows us to derive the most reliable estimates of the AGN luminosity and SFR of W2246$-$0526, based on the currently available data. Our analysis builds on the findings of \cite{varn24}, who conducted a similar study with SMART of the galaxy COS-87259 at $z\sim6.853$, discovered in the Cosmological Evolution Survey (COSMOS) field, using the four combinations of torus models discussed in this paper.

This paper is organized as follows: In Section \ref{sec:data} we provide an overview of the data, while in Section \ref{sec:method} we describe the models and the SED fitting method. In Section \ref{sec:results} we present our results, followed by a discussion in Section \ref{sec:discussion}. Finally, in Section \ref{sec:conclusion} we summarize our conclusions. Throughout this work, we assume $H_0=70$\,km\,s$^{-1}$\,Mpc$^{-1}$, $\Omega=1$ and $\Omega_{\Lambda}=0.7$.

\section{Description of the data}\label{sec:data}

Mid-Infrared Instrument (MIRI) observations of W2246$-$0526 were carried out on Jul 17, 2023 12:13:28 $-$ Jul 17, 2023 15:03:41 (PI: D{\'\i}az-Santos; ID 1712),  using the medium-resolution (MRS) integral field unit (IFU) mode. The instrument was set up to observe the entire wavelength range ($\sim5-28~\mu m$), using the short, medium and long sub-band channels, using a 4-point dither pattern. The readout used for the detector was SLOWR1 with 31 non-destructive readouts for the short grating setup, 23 for the medium grating and 15 for the long grating. The total exposure times where 740.58, 549.47 and 358.35 seconds, respectively. A background/sky observation was taken after the source, offset to an empty patch in the sky free of mid-infrared sources, so that W2246$-$0526 would land on the MIRI imager detector. The readout and exposure time were the same as for the source, but only a 2-dither pattern was used, enough to remove cosmic rays and detector artifacts. Simultaneous images were taken in the F560W, F770W and F1000W fitters.

Data reduction was performed using the standard \textit{JWST} pipeline, version 1.17.1 and CRDS version 1321. Stage 1 of the pipeline executes regular infrared reduction steps, producing count rates. Stage 2 performs flux calibration, corrects for fringing effects, applies a wavelength calibration, WCS and performs sky subtraction. Stage 3 extracts the spectra from the 2-dimensional spectral images and rebuilds them into a 3-dimensional cube, using the drizzle algorithm. A master background was constructed using the offset, sky observations, which was subtracted from the science data to remove the thermal background.

The spectrum from each sub-band was extracted from their respective 3-dimensional cubes, using the CRETA module of the Continuum and Feature Extraction (CAFE)\footnote{\url{https://github.com/GOALS-survey/CAFE}} tool \citep{tanio25}. A conical aperture growing in size linearly with wavelength, suited for unresolved/point sources, was used to extract the spectra. The reference radius was set to 0.5" at a reference wavelength of 5.4 $\mu m$. Since the aperture is larger than the FWHM of the point spread function (PSF) of the instrument at each wavelength, no aperture correction was used. No local background subtraction was applied either. The spectra of the sub-bands were stitched to each other, starting from the longest wavelength sub-band and proceeding in order through shorter wavelength sub-bands.

We reduced the resolution of the \textit{JWST} data as follows: We only used the data for observed wavelength $5~\mu m\leq\lambda<19~\mu m$, as at longer wavelengths the sensitivity of the MIRI instrument degrades and the data become unreliable due to large uncertainties. We extracted the data for wavelengths separated by steps of 0.05 in the logarithm of rest wavelength to match the model wavelength resolution. We added a point at 18.48 $\mu m$ observed wavelength, which corresponds to the wavelength of the 3.3 $\mu m$ polycyclic aromatic hydrocarbon molecule (PAH) feature, which is nevertheless not detected in emission in the spectrum of the galaxy. The extracted data are listed in Table \ref{tab:jwst}.

We complement the \textit{JWST} data with archival multiwavelength photometry from \cite{tsai18}, as listed in Table \ref{tab:data}. These measurements combine space- and ground-based measurements. Optical and near-infrared points come from \textit{HST}/WFC3 (F160W), Keck/OSIRIS (\textit{K} band) and \textit{Spitzer}/IRAC (3.6 and 4.5 $\mu m$). The mid-infrared continuum is sampled by \textit{WISE} at 12 and 22 $\mu m$, while the far-infrared emission is constrained by \textit{Herschel}/Photodetector Array Camera and Spectrometer (PACS) (70 and 160 $\mu m$) and SPIRE ($250-350~\mu m$). Longer-wavelength coverage includes JCMT/SCUBA-2 photometry at 450 and 850 $\mu m$, complemented by high-resolution Atacama Large Millimeter/submillimeter Array (ALMA) data at 882 $\mu m$.

\begin{table}
	\centering
\caption{\textit{JWST}/MIRI-MRS data of W2246$-$0526 used in this work}
	\label{tab:jwst}
	\begin{tabular}{lll} % three columns, alignment for each
		\hline
		Wavelength  &  Flux density &  Error \\
         $\mu m$    &  $\mu$Jy &  $\mu$Jy        \\
		\hline 
5.00  & 105.01 &  20.07 \\
5.60  & 92.01 &  25.01 \\
6.28  & 220.31 &  28.71 \\
7.05  & 129.79 &  39.50 \\
7.91  & 472.29 &  19.70 \\
8.88  & 950.98 &  18.73 \\
9.96  & 1845.45 &  27.68 \\
11.18  & 3152.07 &  43.86 \\
12.54  &  4903.15 &  21.13 \\
14.07  &  7588.54 &  31.13 \\
15.79  &  9982.05 &  37.46 \\ 
17.71  &  13157.08 &  67.27 \\ 
18.48  &  13997.15  &  71.55 \\ 
		\hline
	\end{tabular}
\end{table}

\begin{table}
	\centering
\caption{Photometry of W2246$-$0526 used in this work, as listed by Tsai et al. (2018)}
	\label{tab:data}
	\begin{tabular}{llll} % four columns, alignment for each
		\hline
		Wavelength  &  Flux density &  Error &  Band \\
         $\mu m$    &  $\mu$Jy &  $\mu$Jy &         \\
		\hline 
1.537 &  6.1 &  0.2  & WFC3 F160W \\
2.159   & 8.9  & 2.8    &   \textit{K} band \\
3.6  & 38 &  2   &  IRAC band 1 \\
4.5  & 33 &  1   &  IRAC band 2 \\
12  &  2500 &  200   & \textit{WISE} band 3 \\ 
22  &  15900 &  1600  & \textit{WISE} band 4 \\
70  &  37000 &  3000  &  PACS blue channel \\
160  &  142000 &  16000  & PACS red channel \\
250  &  107000 &  8000  & SPIRE 250 $\mu m$ \\
350  &  81000 &  12000  & SPIRE 350 $\mu m$ \\
450  &  49000 &  12000  & SCUBA-2 450 $\mu m$ \\
850 &   11000 &  2000 & SCUBA-2 850 $\mu m$ \\
882 &  7400 &  400 & ALMA 882 $\mu m$ \\
		\hline
	\end{tabular}
\end{table}

\section{Description of the models and the SED fitting method}\label{sec:method}

Our method, thoroughly detailed in \cite{smart}, allows us to explore the impact of four different AGN torus models and therefore constrain the properties of the obscuring torus, but also quantify the uncertainties in the AGN fraction, black hole mass and SFR of the fitted galaxy. In this study, we explore the option of incorporating a component of polar dust in the fitting. The starburst model of \cite{efstathiou00}, as revised by \cite{efstathiou09}, and the spheroidal host model of \cite{efs21} are fitted in combination with each torus model:

\begin{enumerate}

\item The smooth AGN torus model originally developed by \cite{efstathiou95}, which is part of the CYGNUS (CYprus models for Galaxies and their NUclear Spectra) collection of radiative transfer models. This model assumes a tapered disc geometry (the thickness of the disc increases linearly with distance from the black hole in the inner part of the torus, but assumes a constant thickness in the outer part).

\item The smooth AGN torus model of \cite{fritz06}, which assumes a flared disc geometry (the thickness of the disc increases linearly with distance from the black hole).

\item The two-phase AGN torus model SKIRTOR of \cite{stal16}, which also assumes a flared disc geometry.

\item The two-phase AGN torus model of \cite{sieb15}, which assumes that dust covers the whole sphere around the black hole, i.e. the half-opening angle of the torus is assumed to be zero. Additionally, unlike the other three torus models listed above, this model assumes that dust grains are fluffy, resulting in higher emissivity in the far-infrared and submillimetre. Consequently, this model is expected to exhibit the strongest contribution from the AGN in that part of the spectrum. 

\end{enumerate}

Each AGN torus model has four free parameters. The SKIRTOR and \cite{fritz06} models include two additional parameters that we have fixed. As noted in \cite{smart}, \cite{varn24} and \cite{varn25}, fixing the parameter $p$ of the SKIRTOR model to the value of 1, results in the best agreement for the silicate absorption features of obscured quasars. The parameter $q$ in SKIRTOR concerns the azimuthal dependence of the density distribution and is unlikely to have much impact on the silicate features. As in \cite{smart}, \cite{varn24} and \cite{varn25}, we have fixed this value at 1. Similarly, we have fixed two of the parameters of the \cite{fritz06} model, as detailed in Table \ref{tab:parameters}.

The starburst model has three parameters, which are the initial optical depth of the molecular clouds that constitute the starburst, the e-folding time of the exponential star formation history (SFH) and its age. The spheroidal model also has three parameters, which are the e-folding time of the delayed exponential SFH, the optical depth of the galaxy and the intensity of starlight. The dust and star distribution in the spheroidal galaxy is assumed to have a S\'ersic profile with $n=4$. The parameters of the starburst and spheroidal models are also listed in Table \ref{tab:parameters}.

We have further developed SMART to optionally take into account the extinction from the host galaxy. We used this version of SMART to explore whether incorporating the extinction from the host galaxy can improve the fit to the mid-infrared SED with either of the AGN torus models. The optical depth of the spheroidal host from its center to its surface is one of the model parameters. We can therefore self-consistently apply an extinction to the spectrum of the AGN, assuming the AGN lies at the center of the system.  The wavelength dependence of extinction is assumed to be that of the dust model that is used for the radiative transfer calculations of the starburst and spheroidal components.

The polar dust model assumes that polar dust is concentrated in discrete spherical optically thick clouds, all of which are assumed to have constant temperature for all dust grains \citep{efstathiou06}. The model assumes the same multigrain dust mixture as in the starburst radiative transfer model \citep{efstathiou09}, but it is assumed that the small grains and PAHs are destroyed by the strong radiation field of the AGN to which these clouds are directly exposed. In contrast to the approach of \cite{efs22}, SMART treats the temperature of the polar dust clouds $T_p$ as a free parameter in the fit, which is assumed to vary in the range 800$-$1200 K. Additionally, we assume that all clouds have an optical depth from the center to the surface in the $V$ band of 100. A full radiative transfer calculation in spherical symmetry is carried out for each cloud. Due to the assumed temperature range, polar dust contributes mainly in the near-infrared.

W2246$-$0526 is a merger system (e.g. \citealt{tanio18}) and is therefore more similar to local ULIRGs than quasi stellar objects (QSOs). Although there are broad emission lines observed in this object, it is believed they are seen through patchy obscuration in the AGN torus \citep{tanio16,tsai18,fan20}. In the local Universe, we have examples of hyperluminous obscured AGN that show deep silicate absorption features, such as IRAS~08572+3915 and IRAS~00397-1312 \citep{efs22,smart,varn25}. These objects do not have a luminous polar dust component, so the silicate absorption feature is very deep. In our model, polar dust is located in the region not covered by the torus (see Fig. \ref{fig:polar}).

Notably, the CAT3D-WIND torus model of \cite{hk17} is particularly well matched to the geometry we propose in this paper for W2246$-$0526, but it is not currently implemented in SMART. Its inclusion is planned for future work.

\begin{table*}
	\centering
\caption{Parameters of the models used in this paper, symbols used, their assumed ranges and summary of additional information about the models. The Fritz et al. (2006) model has two additional parameters that define the density distribution in the radial direction ($\beta$) and azimuthal direction ($\gamma$). In this paper, we assume $\beta=0$ and $\gamma=4$. The SKIRTOR model has two additional parameters that define the density distribution in the radial direction ($p$) and azimuthal direction ($q$). In this paper, we assume $p=1$ and $q=1$. In addition, the SKIRTOR library fixes the fraction of mass inside clumps to 97 per cent. There are three additional scaling parameters for the starburst, spheroidal host, AGN torus and polar dust models, $f_{SB}$, $f_{sph}$, $f_{AGN}$ and $f_{p}$, respectively.}
	\label{tab:parameters}
 \resizebox{\textwidth}{!}{\begin{tabular}{llll} % four columns, alignment for each
 \hline
		Parameter &  Symbol & Range &  Comments\\
		\hline
                 &  &  & \\
{\bf CYGNUS Starburst}  &  &  & \\
                 &  &  \\
Initial optical depth of giant molecular clouds & $\tau_v$  &  50$-$250  &  \cite{efstathiou00}, \cite{efstathiou09} \\
Starburst SFR e-folding time       & $\tau_{*}$  & 10$-$35 Myr  & Incorporates \cite{bru93,bru03}  \\
Starburst age      & $t_{*}$   &  5$-$35 Myr &  Metallicity=solar, Salpeter IMF \\ 
                  &            &  & Standard galactic dust mixture with PAHs\\
                  &            &  &  \\                          

                               {\bf CYGNUS Spheroidal Host}  &  &  &  \\
                 &  &  \\
Spheroidal SFR e-folding time      & $\tau^s$  &  0.125$-$8 Gyr  & \cite{efs03}, \cite{efs21}  \\
Starlight intensity      & $\psi^s$ &  1$-$17 &  Incorporates \cite{bru93,bru03} \\ 
Optical depth     & $\tau_{v}^s$ & 0.1$-$15 &  Range of metallicities, Salpeter IMF\\ 
                  &            &  & Standard galactic dust mixture with PAHs \\
                  &            &  &  \\        
                  
{\bf CYGNUS AGN torus}  &  &    &  \\
                 &  &  &  \\
Torus equatorial UV optical depth   & $\tau_{uv}$  &  260$-$1490 &  Smooth tapered discs\\  
Torus ratio of outer to inner radius & $r_2/r_1$ &  20$-$100 & \cite{efstathiou95}, \cite{efstathiou13} \\   
Torus half-opening angle  & $\theta_o$  &  30\degr$-$75\degr & Standard galactic dust mixture without PAHs\\ 
Torus inclination     & $\theta_i$  &  0\degr$-$90\degr &  The subranges $\theta_o$$-$90\degr \hspace{1pt} and 0\degr$-$$\theta_o$ are assumed for\\ 
                  &            &  &  AGN\_type=2 and AGN\_type=1, respectively. \\ 
                 &            & \\
{\bf \cite{fritz06} AGN torus}  &  &  &   \\
                 &  &  & \\
Torus equatorial optical depth at 9.7 $\mu m$  & $\tau_{9.7\mu m}$ &  1$-$10 & Smooth flared discs \\  
Torus ratio of outer to inner radius & $r_2/r_1$ &  10$-$150 &  \cite{fritz06}\\   
Torus half-opening angle  & $\theta_o$  &  20\degr$-$70\degr & Standard galactic dust mixture without PAHs\\ 
Torus inclination     &  $\theta_i$ &  0\degr$-$90\degr &  The subranges $\theta_o$$-$90\degr \hspace{1pt} and 0\degr$-$$\theta_o$ are assumed for\\ 
                  &            &  &  AGN\_type=2 and AGN\_type=1, respectively. \\    
                 &            & &  \\
{\bf SKIRTOR AGN torus}  &  &   &  \\
                 &  &  &  \\
Torus equatorial optical depth at 9.7 $\mu m$  &  $\tau_{9.7\mu m}$ &  3$-$11 & Two-phase flared discs \\  
Torus ratio of outer to inner radius & $r_2/r_1$ &  10$-$30 &  \cite{sta12,stal16} \\   
Torus half-opening angle  & $\theta_o$ &  20\degr$-$70\degr &  Standard galactic dust mixture without PAHs\\ 
Torus inclination     &  $\theta_i$ &  0\degr$-$90\degr & The subranges $\theta_o$$-$90\degr \hspace{1pt} and 0\degr$-$$\theta_o$ are assumed for\\ 
                  &            &  & AGN\_type=2 and AGN\_type=1, respectively. \\   
                 &            & &  \\
{\bf \cite{sieb15} AGN torus}  &  &   &  \\
                 &  &  &  \\
Cloud volume filling factor (per cent)   & $V_c$ &  1.5$-$77  & Two-phase anisotropic spheres \\  
Optical depth of the individual clouds & $A_c$  &  0$-$45 & \cite{sieb15}\\
Optical depth of the disc mid-plane & $A_d$  &  50$-$500 &  Fluffy dust mixture without PAHs\\ 
Inclination     &  $\theta_i$ &   0\degr$-$90\degr & The subranges 45\degr$-$90\degr \hspace{1pt} and 0\degr$-$45\degr are assumed for\\ 
                  &            &  &  AGN\_type=2 and AGN\_type=1, respectively. \\
                                   &            & &  \\
{\bf Polar dust}  &  &   &  \\
                 &  &  &  \\
Temperature  & $T_p$ &  800$-$1200 K  &  Optically thick spherical clouds \citep{efstathiou06} \\  
                  &            &  &   \\                     
		\hline
	\end{tabular}}
\end{table*}

SMART can extract not only the fitted model parameters but also a number of other physical quantities. In Table \ref{tab:derived} we list all the derived physical quantities for W2246$-$0526. We have a total of 13 free parameters in the fits without the polar dust model and 15 free parameters in the fits with the polar dust model, including the scaling factors for each of the components. Table \ref{tab:fitted} gives the selected fitted parameters, along with their errors. The physical quantities we extracted are listed in Tables \ref{tab:extractedA} and \ref{tab:extractedB}.

The SFR of the starburst and the spheroidal component are computed self-consistently by the radiative transfer models, which incorporate the stellar population synthesis models of \cite{bru93,bru03}. We assume a Salpeter initial mass function (IMF) with a metallicity 6.5 per cent of solar for the spheroidal model and solar metallicity for the starburst model. 

In Table \ref{tab:extractedA} we list the SFR of the starburst averaged over the age of the starburst, $\dot{M}^{age}_{*}$, where the age is determined from the fit.

The anisotropic emission from the torus, which is a feature of all torus models considered in this study, requires a correction to the observed luminosity of the AGN to determine their intrinsic luminosity. The anisotropy correction factor $A$, as defined by \cite{efstathiou06}, \cite{efs22}, \cite{smart}, \cite{varn24} and \cite{varn25}, represents the factor by which we need to multiply the observed luminosity to get the true luminosity:
\begin{equation}
A(\theta_i)={{\int_0^{\pi/2}S(\theta_i')~sin\theta_i'~d\theta_i' }\over{S(\theta_i)}}~~,
\end{equation}
where $\theta_i$ is the torus inclination and $S(\theta_i)$ is the emission integrated over the relevant wavelength range. $A(\theta_i)$ varies for the infrared and bolometric luminosities. In Table \ref{tab:extractedB} we list the anisotropy correction factor $A$ predicted by the four different torus models, with and without polar dust.

For this study, SMART offers insights into the properties of W2246$-$0526 that grid-based methods, such as Code Investigating GALaxy Emission (CIGALE; \citealt{noll09,boq19}), cannot provide. \textit{JWST}/MIRI data contain features from dust, including PAH and silicate features, that are particularly important for constraining the energy source of galaxies. Grid-based approaches restrict model parameters to discrete values and this limits the ability to model highly sampled spectrophotometry data, which are essential for decomposing the SED of W2246$-$0526 and inferring the contribution of star formation and AGN activity. Our method instead explores the full parameter space, which is especially critical for key parameters like the AGN torus inclination, which significantly affects the luminosity of the system. Furthermore, by employing radiative transfer models, we take properly into account the effects of dust in a realistic geometry, whereas energy-balance methods, such as Multi-wavelength Analysis of Galaxy Physical Properties (MAGPHYS; \citealt{dac08}) and CIGALE, are limited by the implementation of attenuation laws, which may affect the estimates of stellar mass and SFR.

\begin{figure*}
	\begin{center}
      {\includegraphics[width=85mm]{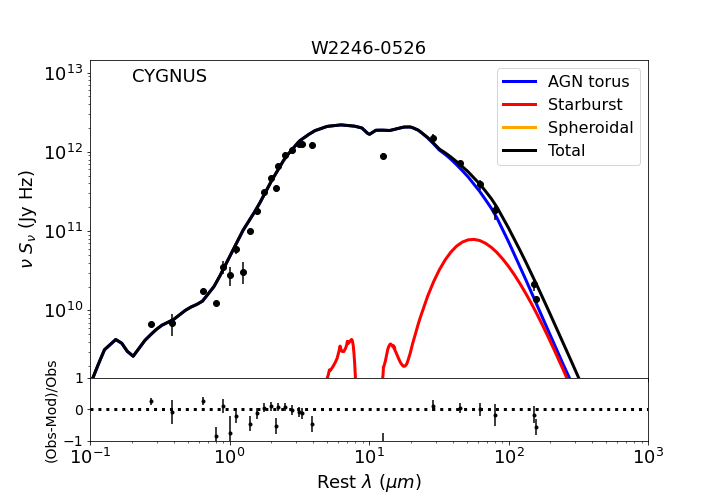}}    
      {\includegraphics[width=85mm]{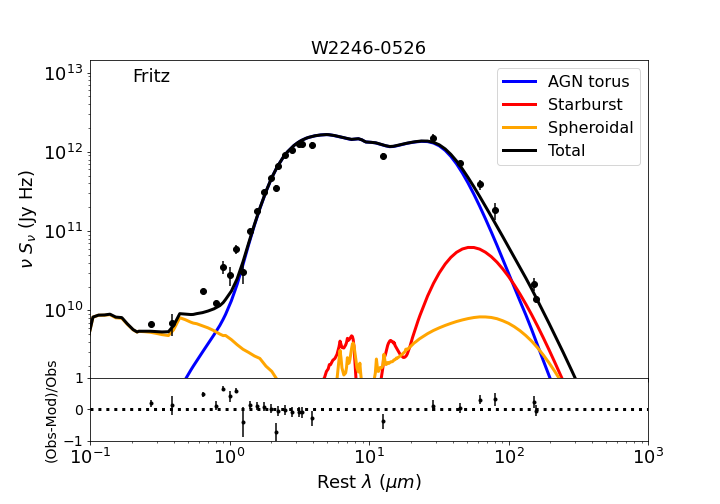}}     
      {\includegraphics[width=85mm]{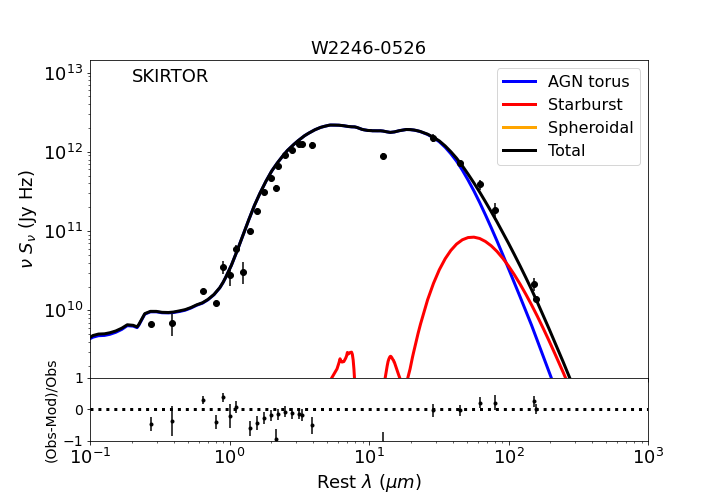}}   
      {\includegraphics[width=85mm]{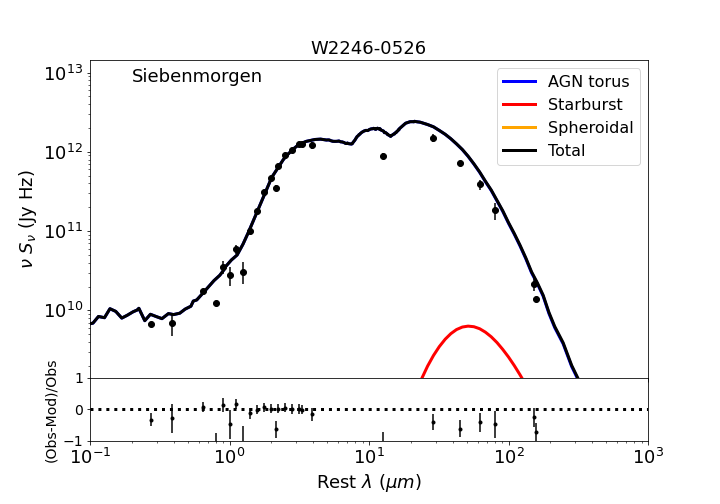}} 
\caption{Comparison SED fit plots of W2246$-$0526 without the addition of host extiction or the polar dust component. The AGN torus, starburst, spheroidal host and total emissions are plotted as shown in the legend. The top left panel shows fits with the CYGNUS combination of models. The top right panel shows fits with the CYGNUS AGN torus model replaced by the Fritz et al. (2006) model, the bottom left panel replaces the CYGNUS AGN torus model with the SKIRTOR model, while the bottom right panel replaces the CYGNUS AGN torus model with the Siebenmorgen et al. (2015) model.}
\label{fig:4AGN_no_polar}      
      \end{center}
\end{figure*}

\begin{figure*}
	\begin{center}
      {\includegraphics[width=85mm]{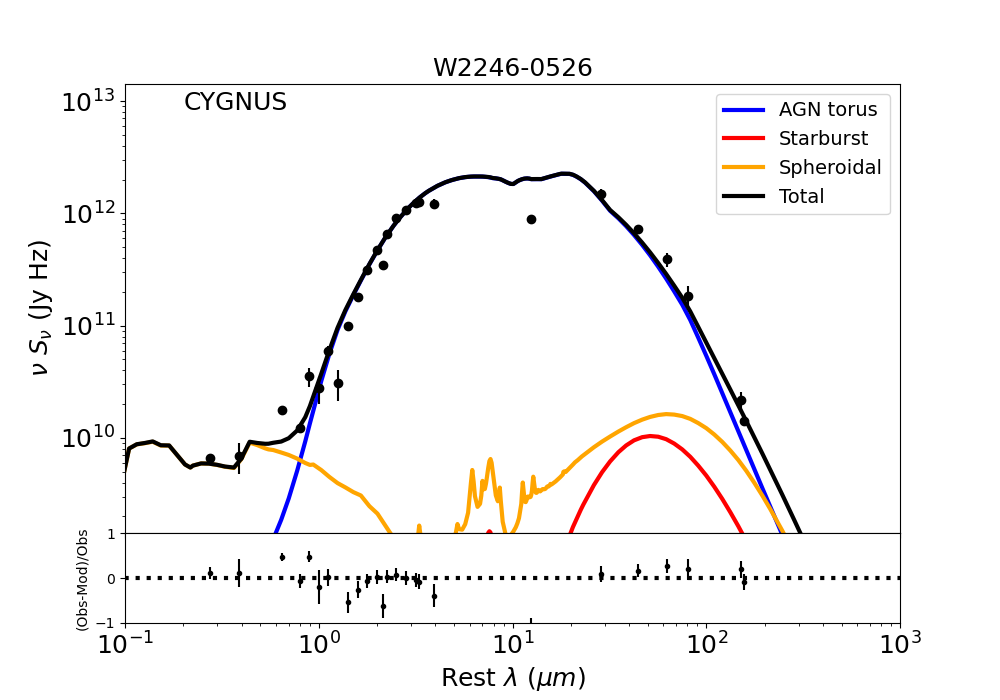}}    
      {\includegraphics[width=85mm]{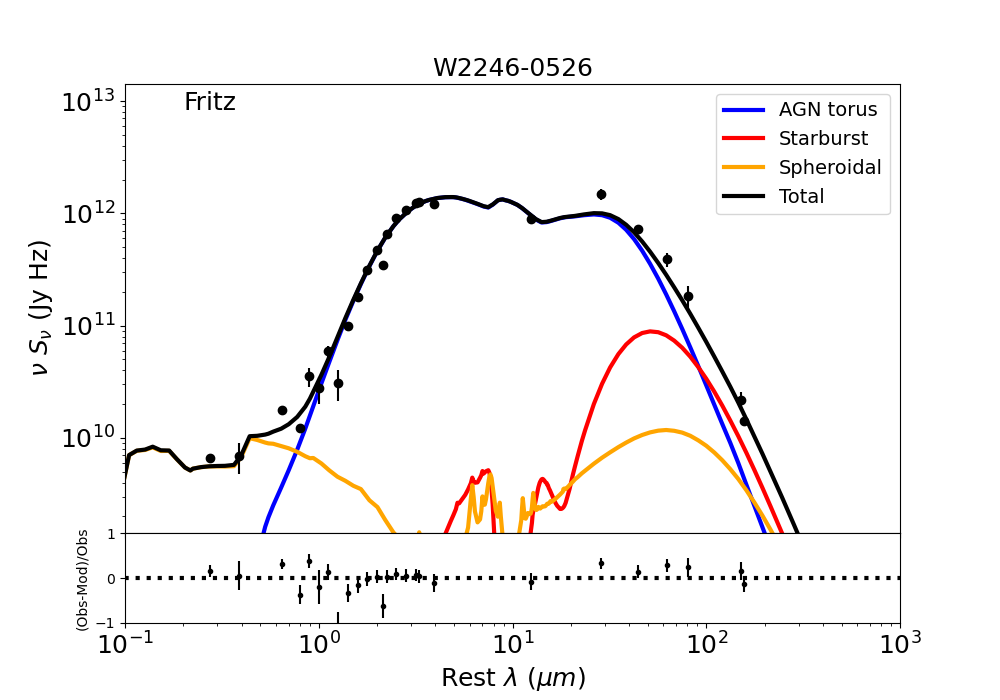}}     
      {\includegraphics[width=85mm]{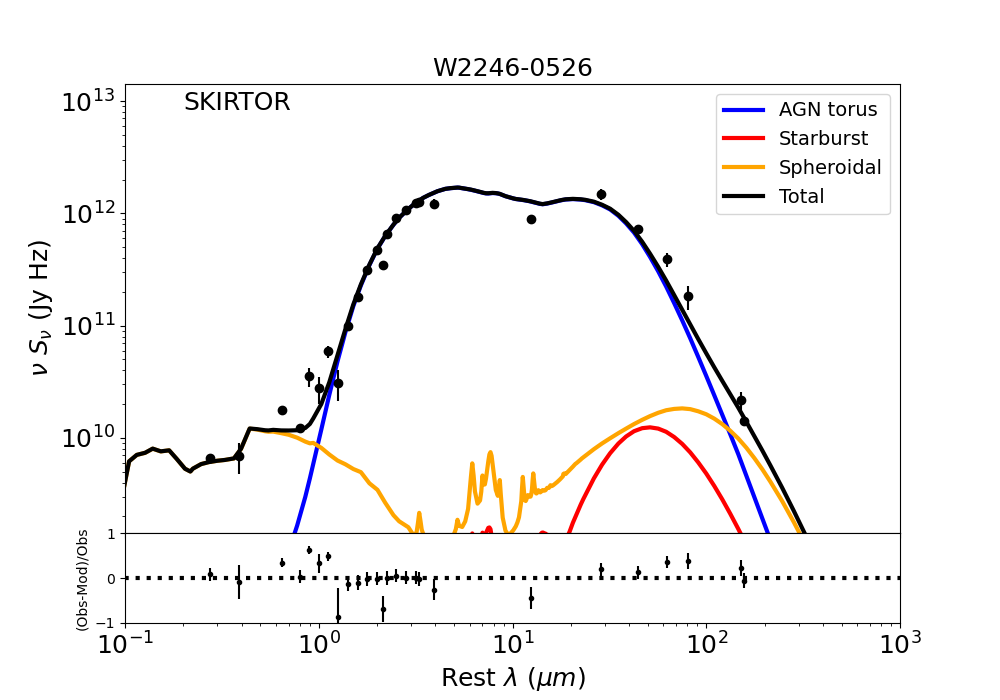}}   
      {\includegraphics[width=85mm]{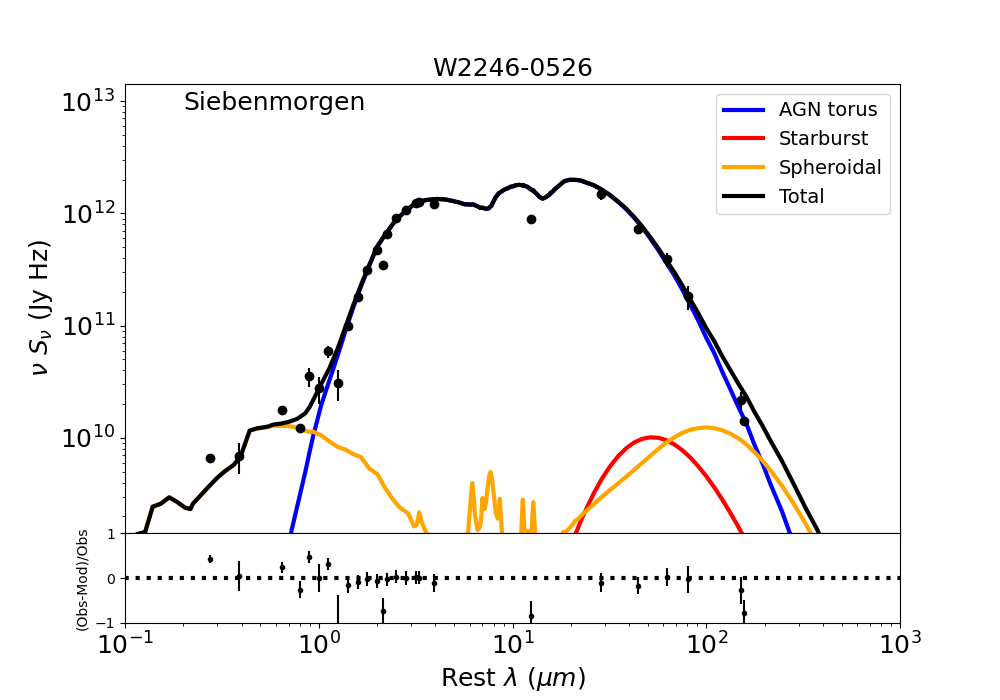}} 
\caption{Comparison SED fit plots of W2246$-$0526 with the addition of host extinction. The AGN torus, starburst, spheroidal host and total emissions are plotted as shown in the legend. The top left panel shows fits with the CYGNUS combination of models. The top right panel shows fits with the CYGNUS AGN torus model replaced by the Fritz et al. (2006) model, the bottom left panel replaces the CYGNUS AGN torus model with the SKIRTOR model, while the bottom right panel replaces the CYGNUS AGN torus model with the Siebenmorgen et al. (2015) model.}
\label{fig:4AGN_extinction}      
      \end{center}
\end{figure*}

\begin{figure*}
	\begin{center}
      {\includegraphics[width=85mm]{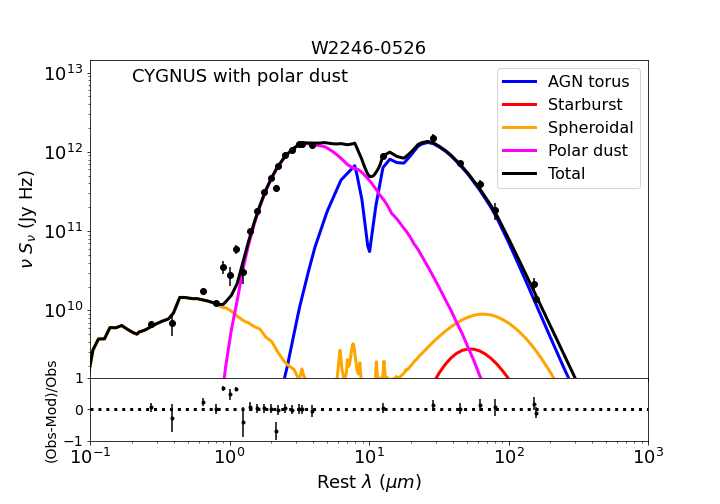}}    
      {\includegraphics[width=85mm]{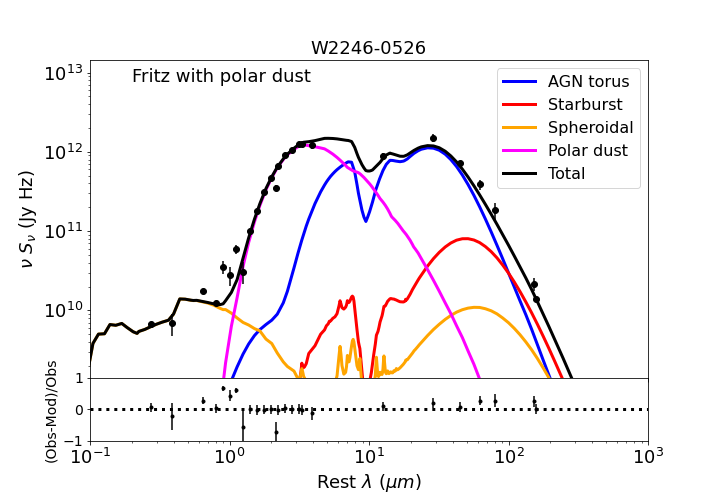}}    
      {\includegraphics[width=85mm]{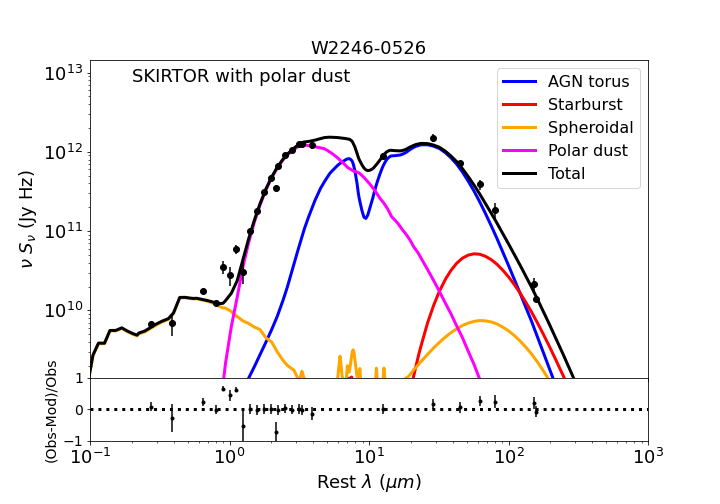}} 
\caption{Comparison SED fit plots of W2246$-$0526 with the addition of the polar dust component. The AGN torus, starburst, spheroidal host, polar dust and total emissions are plotted as shown in the legend. The top left panel shows fits with the CYGNUS combination of models. The top right panel shows fits with the CYGNUS AGN torus model replaced by the Fritz et al. (2006) model and the bottom panel replaces the CYGNUS AGN torus model with the SKIRTOR model.}
\label{fig:4AGN_polar}      
      \end{center}
\end{figure*}

\begin{table}
	\centering
\caption{Derived physical quantities and the symbol used. All luminosities are integrated over $1-1000~\mu m$.}
	\label{tab:derived}
	\begin{tabular}{llll} % two columns, alignment for each
		\hline
		Physical quantity &  Symbol \\
		\hline  
Observed AGN torus luminosity              & $L_{AGN}^{o}$    \\ 
Corrected AGN torus luminosity             & $L_{AGN}^{c}$  \\ 
Starburst luminosity                   & $L_{SB}$  \\
Spheroidal host luminosity             & $L_{sph}$  \\
Polar dust AGN luminosity              & $L_{p}$ \\  
Total corrected luminosity             & $L_{tot}^{c}$  \\ 
Starburst SFR (averaged over its age)   & $\dot{M}_*^{age}$  \\
Spheroidal SFR                         & $\dot{M}_{sph}$  \\
Total SFR                              & $\dot{M}_{tot}$  \\
Spheroidal stellar mass                & $M^{*}_{sph}$  \\ 
Starburst stellar mass                 & $M^{*}_{SB}$  \\ 
Total stellar mass                     & ${M}^{*}_{tot}$  \\ 
AGN fraction                           & $F_{AGN}$  \\
Anisotropy correction factor           & $A$  \\ 
Covering factor of polar dust           & $f_c$  \\ 
Estimated black hole mass              & $M_{BH}$  \\
		\hline
	\end{tabular}
\end{table}

\section{Results}\label{sec:results}

We first discuss the fits without the addition of host extinction or polar dust, which are presented in Fig. \ref{fig:4AGN_no_polar}. These fits show a number of interesting characteristics. First of all, we see that the solutions for the CYGNUS, SKIRTOR and \cite{sieb15} models fit the rest frame optical emission with torus emission, instead of the spheroidal component. In the cases of the CYGNUS and SKIRTOR models, the predicted inclination is just greater than the half-opening angle of the torus, which explains this effect as we are seeing the hot other side of the torus and possibly scattered light. In the case of the \cite{sieb15} model, the half-opening angle of the torus is by definition 0. In the \cite{fritz06} model fit, the optical/ultraviolet spectrum is fitted with the spheroidal component. Another important characteristic is that the fits with all models fail to fit satisfactorily the PACS 70 $\mu m$ point and tend to predict relatively low optical depths for the torus. In all cases, there is only a very weak contribution in the far-infrared and submillimetre from a starburst or the spheroidal host.

We attribute the failure of `pure' torus models, with the addition of a starburst and a spheroidal host, to explain the complete SED of W2246$-$0526 to the fact that this object simultaneously shows the presence of hot dust and high obscuration, as indicated by the 70 $\mu m$ point. These two features appear to be incompatible, as high obscuration has the effect of absorbing the emission by hot dust in the inner part of the torus. The addition of hot polar dust that suffers less obscuration than the hot inner edge of the torus offers the possibility to better explain the SED. In our method, we fix two of the parameters of the SKIRTOR and \cite{fritz06} models, which have to do with the density distribution in the torus. We do not consider this to be the reason we fail to fit the SED with a `pure' torus model. In the case of the CYGNUS model, we allow all of the parameters to vary. In the case of the \cite{sieb15} model, we allow all the parameters to vary, except the inner radius of the torus that is fixed to give an inner temperature of around 1000K.

The results of the fitting with host extinction included are shown in Fig. \ref{fig:4AGN_extinction}. We find that incorporating the extinction from the host galaxy does not significantly improve the fits. This may be related to the fact that the optical depths of the spheroidal component are relatively low, and are found to be in the range $2.2-5.3$. This low-level of host extinction is not sufficient to suppress the mid-infrared emission. The minimum reduced $\chi^2$ and selected fitted parameters of the fits with host extinction are shown in Table \ref{tab:fitted}.

In the optical/near-infrared, the emission from the spheroidal component is degenerate with scattered light from the AGN, which is predicted to be viewed from a very special inclination (angle just greater than the half-opening angle of the torus). Attempts to enforce a fit with a spheroidal component were unsuccessful. This difficulty motivated our adoption of a torus plus polar dust geometry.

\begin{table}
	\centering
\caption{Results of the $\chi^2$ difference test (described in Appendix \ref{chi_2_diff}) and BIC (described in Appendix \ref{BIC}) for each AGN torus model, with and without the polar dust component}
	\label{tab:test}
 \begin{tabular}{ lcccccccc }
 \hline 
AGN torus model & $\text{p-value}$ & $\Delta\mathrm{BIC}$\\ 
 \hline
 CYGNUS & 0.00003 & -14.18 \\ 
 \\   
  \cite{fritz06} & 0.08521 & 1.59 \\ 
 \\   
 SKIRTOR & 0.00002 & -14.84 \\  
 \hline             
 \end{tabular}
\end{table}

\begin{table*}
	\centering
\caption{Minimum reduced $\chi^2$ and selected fitted parameters for W2246$-$0526. A description of the parameters is given in Table \ref{tab:parameters}. For the CYGNUS model, $\tau_{uv}$ is divided by 61 to obtain $\tau_{9.7\mu m}$.}
	\label{tab:fitted}
 \begin{tabular}{ lcccccccc }
 \hline 
AGN torus model &  $\chi^{2}_{min, \nu}$ & $\tau^s \ (10^7yr)$ & $\tau_v$ & $t_{*} \ (10^7yr)$  & & $\theta_o \ (\degr)$  & $\theta_i \ (\degr)$ & $T_p$ \\ 
 \hline $  $ & & & & & $\tau_{9.7\mu m}$ \\  
 CYGNUS without polar dust & 5.1 & $69.3^{9.8}_{-13.1}$  &    $155.1^{11.1}_{-6.7}$ & $1.7^{0.3}_{-0.5}$ & $4.5^{0.1}_{-0.1}$  &    $63.5^{3.0}_{-0.0}$  &    $65.9^{1.9}_{-0.7}$ & - \\ 
 or host extinction
 \\ \\
 $  $ & & & & & $\tau_{9.7\mu m}$ \\
 CYGNUS with host extinction & 4.8 & $218.2^{32.3}_{-66.5}$  &    $89.8^{13.0}_{-18.5}$ & $1.6^{0.1}_{-0.1}$ & $4.3^{0.1}_{-0.0}$  &    $57.9^{0.1}_{-0.0}$  &    $56.9^{2.2}_{-1.5}$ & - \\ 
 \\ 
 $  $ & & & & & $\tau_{9.7\mu m}$ \\
 CYGNUS with polar dust &  3.2  &  $37.3^{1.7}_{-2.6}$  &    $119.9^{48.1}_{-23.6}$ & $1.4^{0.7}_{-0.4}$ & $9.7^{0.7}_{-1.1}$  &    $53.0^{6.4}_{-4.6}$  &    $76.2^{3.5}_{-5.1}$ &     $940.8^{5.7}_{-2.1}$ \\ 
 \\   
 $  $ & & & & & $\tau_{9.7\mu m}$ \\ 
 \cite{fritz06} without polar dust & 3.9 & $155.2^{11.6}_{-26.6}$  &    $129.2^{3.1}_{-7.4}$ & $1.4^{0.0}_{-0.2}$ & $5.8^{1.0}_{-0.4}$  &    $40.1^{0.4}_{-1.9}$  &    $59.4^{3.2}_{-0.9}$ & - \\ 
 or host extinction
 \\ \\
 $  $ & & & & & $\tau_{9.7\mu m}$ \\ 
 \cite{fritz06} with host extinction & 3.7 & $74.4^{58.8}_{-21.6}$  &    $139.3^{0.7}_{-51.4}$ & $0.7^{0.4}_{-0.0}$ & $3.3^{0.6}_{-0.7}$  &    $45.9^{1.2}_{-0.1}$  &    $66.1^{0.9}_{-2.5}$ & - \\  
 \\   
 $  $ & & & & & $\tau_{9.7\mu m}$ \\ 
 \cite{fritz06} with polar dust & 3.5 & $40.3^{1.8}_{-1.9}$  &    $66.4^{8.2}_{-6.1}$ & $0.9^{0.2}_{-0.2}$ & $8.8^{0.5}_{-0.7}$  &    $35.7^{4.4}_{-3.1}$  &    $83.5^{0.7}_{-0.6}$ &     $957.4^{4.6}_{-3.2}$ \\ 
 \\
 $  $ & & & & & $\tau_{9.7\mu m}$ \\ 
 SKIRTOR without polar dust & 5.3 & $88.2^{51.9}_{-45.3}$  &    $129.3^{35.1}_{-19.8}$ & $1.6^{0.3}_{-0.4}$ & $6.9^{1.5}_{-0.6}$  &    $54.2^{0.3}_{-0.3}$  &    $59.7^{1.6}_{-1.8}$ & - \\  
 or host extinction
 \\ \\
 $  $ & & & & & $\tau_{9.7\mu m}$ \\
 SKIRTOR with host extinction & 3.7 & $51.5^{17.7}_{-5.2}$  &    $89.9^{18.4}_{-12.6}$ & $0.9^{0.2}_{-0.3}$ & $7.6^{1.0}_{-1.0}$  &    $56.9^{0.5}_{-0.2}$  &    $51.0^{3.6}_{-3.5}$ & - \\  
 \\ 
 $  $ & & & & & $\tau_{9.7\mu m}$ \\
 SKIRTOR with polar dust & 3.3 & $35.0^{2.5}_{-2.0}$  &    $199.9^{20.2}_{-20.5}$ & $2.3^{0.4}_{-0.4}$ & $10.7^{0.3}_{-0.4}$  &    $44.9^{6.4}_{-3.3}$  &    $78.9^{1.1}_{-1.5}$ &     $957.1^{3.6}_{-4.3}$ \\  
 \\
 $  $ & & & & & $A_d$ \\ 
 \cite{sieb15} without & 4.9 & $383.2^{77.1}_{-196.8}$  &    $101.6^{20.9}_{-11.3}$ & $1.1^{0.2}_{-0.2}$ & $53.5^{0.8}_{-0.4}$ & 0.0 & $66.8^{0.5}_{-0.5}$ & - \\ 
 host extinction
 \\ \\
  $  $ & & & & & $A_d$ \\ 
 \cite{sieb15} with host & 3.6 & $30.3^{0.4}_{-0.4}$  &    $103.3^{5.3}_{-5.3}$ & $1.9^{0.1}_{-0.1}$ & $52.6^{1.5}_{-1.5}$ & 0.0 & $61.0^{0.3}_{-0.3}$ & - \\ 
 extinction 
 \\ \\
 \hline              
 \end{tabular}
\end{table*}

In Fig. \ref{fig:4AGN_polar} we present the fits of the three models, CYGNUS, \cite{fritz06} and SKIRTOR, obtained if we include polar dust. We note that the \cite{sieb15} model already includes polar dust, so we do not consider a model in which we add polar dust. Our analysis allows us to assess which combination of models best fits the observational data for W2246$-$0526. From the SED fit plots, we observe that the fits with polar dust enabled are better than those without polar dust inclusion for all torus models considered. This is probably a hint that none of the current torus models alone can provide a satisfactory explanation for the near- and mid-infrared spectrum of this quasar. We argue that this may be evidence for polar dust in W2246$-$0526. 

Although the \cite{sieb15} model also includes polar dust in a self-consistent radiative transfer calculation, the model does not have the flexibility of the approach we adopt for the other three models, where we have the temperature of the polar dust clouds as a parameter and also the covering factor or luminosity of the polar dust (through the scaling parameter $f_p$). We also note that radiation from the central source may not be emitted isotropically, as assumed in the \cite{sieb15} model. The \cite{sieb15} model has a more rigid structure, which is determined by the density distribution of the dust around the AGN and the isotropic emission of the central source, and this is probably the main reason why it does not work as well in this case. Given that the two-phase SKIRTOR model with the addition of polar dust gives a good fit, we can speculate that the \cite{sieb15} model would also give a good fit, if the treatment of polar dust were more flexible.

It is interesting that in all three other models we have a similar solution of a torus viewed at a higher inclination (more edge-on) than in the case without polar dust. The torus dominates at the longer wavelengths and produces a deep silicate absorption feature. The polar dust dominates in the near-infrared, where the \textit{JWST} data lie and tend to make the silicate absorption feature much shallower in the total emission. In all three model combinations, we have a small contribution from the starburst and spheroidal components in the far-infrared, with the strongest contribution in the case of the \cite{fritz06} model. However, because of the extreme luminosity of this system, the resulting SFR can approach $3000~M_\odot yr^{-1}$. We also note that in all three cases the rest frame optical/ultraviolet spectrum is dominated by the spheroidal component.

To take into account the additional parameters in the fits with polar dust, in Table \ref{tab:test} we present the results of a $\chi^2$ difference test for nested models and a Bayesian Information Criterion (BIC) for model comparison. For the CYGNUS and SKIRTOR models, including polar dust leads to a statistically significant improvement in the fits ($\text{p}<10^{-4}$). The BIC analysis is consistent with these results, with $\Delta\mathrm{BIC}<-14$ for the CYGNUS and SKIRTOR models, indicating that the improved fits outweigh the penalty for additional parameters and providing very strong evidence for the inclusion of polar dust. In contrast, no significant evidence is found for the \cite{fritz06} model, with $\text{p}>10^{-4}$ and $\Delta\mathrm{BIC}=1.59$.
 
In Table \ref{tab:fitted} we observe that the minimum reduced $\chi^2$ values provided by the fits with polar dust disabled are considerably higher than those obtained when polar dust is enabled, regardless of whether host extinction is included. The fits with polar dust enabled yield similar values across all models. Specifically, the CYGNUS torus model gives the lowest value of $\chi^{2}_{min, \nu}=3.2$, closely followed by the SKIRTOR model with $\chi^{2}_{min, \nu}=3.3$ and the model of \cite{fritz06}, which gives the highest value of $\chi^{2}_{min, \nu}=3.5$. The results suggest that all models, when incorporating polar dust, provide satisfactory fits to the data of W2246$-$0526. 
 
\begin{figure}
\centering
\includegraphics[width=.97\linewidth]{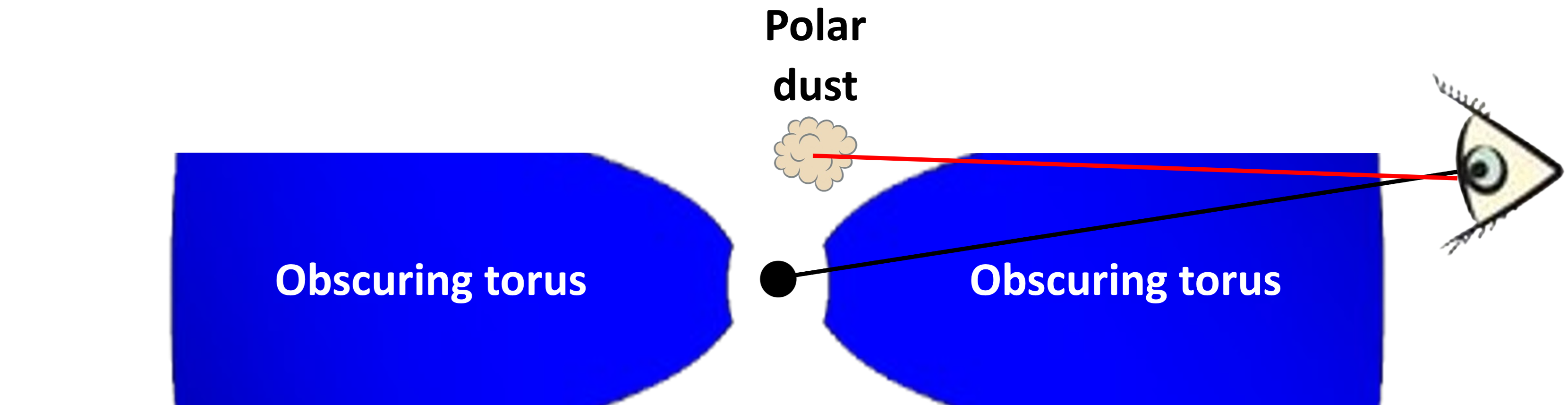}
\caption{Schematic diagram of the possible torus and polar dust geometry that could explain the observed properties of W2246$-$0526. As shown in the diagram, the polar dust emission may suffer some extinction through the torus along the red line of sight.}
\label{fig:polar}
\end{figure}

We conclude that the comparison of the SED fit plots shown in Figs \ref{fig:4AGN_no_polar}, \ref{fig:4AGN_extinction} and \ref{fig:4AGN_polar} is considered indirect evidence for the presence of a component of near- to mid-infrared emission associated with the polar region. For all models, the inclusion of polar dust statistically significantly improves their fit to the data of W2246$-$0526. Considering all the above, we argue that the observed infrared SED of W2246$-$0526 can be most plausibly explained by re-radiation by optically thick dust clouds in the polar regions of the torus, as well as an optically thick torus viewed almost edge-on. In Fig. \ref{fig:polar} we show a schematic diagram of the possible torus and polar dust geometry that could explain the observed properties of W2246$-$0526. Since the models with polar dust enabled perform considerably better, we consider only the results of these combinations in the remainder of this discussion.

According to the fits, the luminosity of this object is dominated by the AGN with an infrared ($1-1000~\mu m$) luminosity (including the anisotropy correction) of $4.2-7.2 \times 10^{14} L_\odot$. Our method predicts a significant anisotropy correction factor $A$ in the range $2.2-3.9$. The SKIRTOR model predicts the lowest A and the CYGNUS model the highest. The CYGNUS model also predicts the highest AGN luminosity of $7.2 \times 10^{14} L_\odot$. All models predict a similar polar dust luminosity in the range $1.6-1.7 \times 10^{14} L_\odot$ and a similar temperature in the range 940.8$-$957.4 K. The CYGNUS model predicts a relatively low starburst luminosity of $4.4 \times 10^{11} L_\odot$. The SKIRTOR model predicts a much higher starburst luminosity of $5.9 \times 10^{12} L_\odot$, while the model of \cite{fritz06} predicts the highest starburst luminosity of $1.1 \times 10^{13} L_\odot$. Similarly, the CYGNUS model predicts a low SFR of $358.3~M_\odot yr^{-1}$. In contrast, the SKIRTOR and \cite{fritz06} models predict a high SFR, ranging between $1714-2916~M_\odot yr^{-1}$. 

The stellar mass of the galaxy is predicted to be in the range $4.8-5 \times 10^{11} M_\odot$. The \cite{fritz06} and SKIRTOR models predict that a significant fraction of the stellar mass, ranging between $2.3-2.9 \times 10^{10} M_\odot$, formed in the current starburst episode in the galaxy. The AGN fraction is predicted to be in the range $72-81$ per cent and the covering factor of polar dust in the range $24-38$ per cent. The method predicts a relatively young starburst age of $0.9-2.3 \times 10^7yr$. The black hole mass (assuming the AGN is accreting at the Eddington limit) is estimated to be in the range $1.3-2.3 \times 10^{10} M_\odot$, with the CYGNUS model giving the highest estimate and the SKIRTOR model the lowest. The models predict that the torus half-opening angle is in the range $35.7\degr-53\degr$, with the CYGNUS model predicting the highest value and the model of \cite{fritz06} the lowest. This implies that the torus in obscured quasars like W2246$-$0526 has a large covering factor and these objects can be at least as numerous as unobscured quasars. We note that, in the case of the two-phase SKIRTOR model, the clumpiness of the medium must be taken into account to calculate the covering factor.  

\begin{table*}
	\centering
\caption{Selected extracted physical quantities for W2246$-$0526. A description of the physical quantities is given in Table \ref{tab:derived}. The reported luminosities are integrated over $1-1000~\mu m$.}
	\label{tab:extractedA}
 \begin{tabular}{ lcccccccc }
 \hline 
AGN torus model & $L_{AGN}^o$ & $L_{AGN}^c$ & $L_{SB}$ & $L_{sph}$ & $L_{p}$ & $L_{tot}^c$ & $\dot{M}^{age}_{*}$ & $\dot{M}_{sph}$ \\  
& $10^{14} L_\odot$ & $10^{14} L_\odot$ & $10^{11} L_\odot$ & $10^{12} L_\odot$ & $10^{13} L_\odot$ & $10^{14} L_\odot$ & $M_\odot yr^{-1}$ & $M_\odot yr^{-1}$ \\    
 \hline CYGNUS without polar dust & $5.2^{0.9}_{-0.1}$  &    $8.7^{0.6}_{-0.2}$  &    $90.1^{44.1}_{-63.8}$  &    $0.0^{0.1}_{-0.0}$  & - &   $8.8^{0.6}_{-0.2}$  &    $2166.0^{1090.0}_{-1546.0}$  &    $0.9^{13.4}_{-0.8}$ \\ 
 \\   
 CYGNUS with polar dust & $1.8^{0.2}_{-0.2}$  &    $7.2^{2.3}_{-1.4}$  &    $4.4^{8.8}_{-2.9}$  &    $2.0^{0.3}_{-0.2}$  &    $17.0^{0.1}_{-0.2}$  &     $9.0^{2.2}_{-1.4}$  &    $110.9^{202.4}_{-72.7}$  &    $217.6^{28.2}_{-25.1}$ \\ 
 \\
  \cite{fritz06} without polar dust & $3.9^{0.1}_{-0.3}$  &    $4.1^{0.1}_{-0.2}$  &    $72.4^{63.2}_{-27.2}$  &    $1.8^{0.5}_{-0.2}$  & - &   $4.2^{0.0}_{-0.2}$  &    $1741.0^{1481.0}_{-670.2}$  &    $259.5^{51.1}_{-39.7}$ \\ 
 \\   
 \cite{fritz06} with polar dust & $1.8^{0.1}_{-0.1}$  &    $4.5^{0.4}_{-0.3}$  &    $110.1^{9.7}_{-12.9}$  &    $2.2^{0.2}_{-0.2}$  &    $16.2^{0.2}_{-0.3}$  &     $6.2^{0.4}_{-0.2}$  &    $2684.0^{385.5}_{-257.8}$  &    $241.0^{19.4}_{-23.8}$ \\ 
 \\
 SKIRTOR without polar dust & $4.6^{0.2}_{-0.2}$  &    $6.1^{0.4}_{-0.3}$  &    $94.0^{19.1}_{-21.8}$  &    $0.0^{0.1}_{-0.0}$ & - &    $6.2^{0.4}_{-0.3}$  &    $2374.0^{369.8}_{-540.6}$  &    $1.8^{13.2}_{-1.7}$ \\  
 \\   
 SKIRTOR with polar dust & $2.0^{0.2}_{-0.2}$  &    $4.2^{0.7}_{-0.3}$  &    $58.6^{6.4}_{-5.9}$  &    $1.7^{0.2}_{-0.2}$  &    $16.1^{0.1}_{-0.3}$  &     $5.9^{0.7}_{-0.3}$  &    $1497.0^{102.9}_{-137.4}$  &    $189.8^{23.6}_{-24.7}$ \\  
 \\   
 \cite{sieb15} & $5.4^{0.1}_{-0.1}$  &    $7.9^{0.1}_{-0.1}$  &    $6.7^{5.9}_{-1.9}$  &    $0.0^{0.0}_{-0.0}$ & - &    $7.9^{0.2}_{-0.1}$  &    $158.5^{138.1}_{-48.6}$  &    $0.3^{0.4}_{-0.2}$ \\ 
 \hline               
 \end{tabular}
\end{table*}

\begin{table*}
	\centering
\caption{Other extracted physical quantities for W2246$-$052. A description of the physical quantities is given in Table \ref{tab:derived}.}
	\label{tab:extractedB}
 \begin{tabular}{ lcccccccc }
 \hline 
AGN torus model & $\dot{M}_{tot}$ & ${M}^{*}_{sph}$ & ${M}^{*}_{SB}$ & ${M}^{*}_{tot}$ & $F_{AGN}$ & $A$ & $f_c$ & $M_{BH}$ \\ 
     & $M_\odot yr^{-1}$ & $10^{11} M_\odot$ & $10^{9} M_\odot$ & $10^{11} M_\odot$ & $   $ & $   $ & & $10^{10} M_\odot$ \\  
 \hline CYGNUS without polar dust & $2167.0^{1091.0}_{-1532.0}$  &    $0.0^{0.1}_{-0.0}$  &    $32.8^{23.3}_{-24.7}$  &    $0.3^{0.2}_{-0.1}$  &     $0.99^{0.01}_{-0.01}$  &     $1.7^{0.0}_{-0.3}$ & - & $2.7^{0.2}_{-0.1}$ \\  
\\
 CYGNUS with polar dust & $358.3^{164.9}_{-109.5}$  &    $4.9^{0.2}_{-0.3}$  &    $1.0^{2.5}_{-0.3}$  &    $4.9^{0.2}_{-0.3}$  &     $0.81^{0.04}_{-0.04}$  &     $3.9^{1.0}_{-0.5}$ & $0.24^{0.08}_{-0.05}$
 & $2.3^{0.6}_{-0.5}$ \\  
\\
 \cite{fritz06} without polar dust & $2001.0^{1441.0}_{-619.0}$  &    $1.6^{0.4}_{-0.0}$  &    $23.0^{17.8}_{-12.1}$  &    $2.0^{0.1}_{-0.1}$  &     $0.98^{0.01}_{-0.01}$  &     $1.1^{0.1}_{-0.1}$ & - & $1.2^{0.0}_{-0.0}$\\ 
\\   
 \cite{fritz06} with polar dust & $2916.0^{401.6}_{-242.7}$  &    $4.6^{0.2}_{-0.4}$  &    $23.4^{4.0}_{-2.6}$  &    $4.8^{0.2}_{-0.3}$  &     $0.72^{0.02}_{-0.02}$  &     $2.5^{0.1}_{-0.1}$ & $0.36^{0.03}_{-0.02}$ & $1.4^{0.1}_{-0.1}$ \\ 
\\   
 SKIRTOR without polar dust & $2383.0^{360.9}_{-549.4}$  &    $0.0^{0.1}_{-0.0}$  &    $30.8^{11.7}_{-3.8}$  &    $0.3^{0.2}_{-0.1}$  &     $0.98^{0.0}_{-0.0}$  &     $1.3^{0.1}_{-0.0}$ & - & $1.9^{0.2}_{-0.1}$ \\
\\  
 SKIRTOR with polar dust & $1714.0^{63.6}_{-186.5}$  &    $4.7^{0.3}_{-0.3}$  &    $29.1^{8.1}_{-4.1}$  &    $5.0^{0.3}_{-0.3}$  &     $0.72^{0.03}_{-0.01}$  &     $2.2^{0.2}_{-0.3}$ & $0.38^{0.06}_{-0.03}$ & $1.3^{0.2}_{-0.1}$ \\
\\   
 \cite{sieb15} & $159.0^{137.9}_{-49.0}$  &    $0.0^{0.0}_{-0.0}$  &    $2.0^{0.8}_{-0.7}$  &    $0.0^{0.0}_{-0.0}$  &     $1.0^{0.0}_{-0.0}$  &     $1.5^{0.0}_{-0.0}$ & - & $2.4^{0.1}_{-0.0}$ \\
 \hline                
 \end{tabular}
\end{table*}

A common characteristic of all the fits without polar dust is that the AGN torus models predict a spectrum with a very shallow silicate absorption feature, if any. This is unexpected, considering that W2246$-$0526 is classified as an obscured quasar. There are examples of local ULIRGs that exhibit very deep silicate absorption features, such as IRAS~08572+3915 \citep{efs14,efs22,smart,varn25}. The shallow absorption feature can be attributed to the fact that the AGN torus models do not predict a very edge-on inclination. Interestingly, when polar dust is incorporated, the models predict higher inclinations and produce a spectrum with a deeper silicate absorption feature, aligning better with what would be expected for obscured quasars. This occurs because the addition of polar dust allows solutions where the AGN torus is more edge-on compared with the case where the effect of polar dust is not taken into account.

Another interesting result is that all fits that include polar dust show the optical/ultraviolet emission to be dominated by the host galaxy, whereas in the fits without polar dust it is dominated by the AGN, with the exception of the model of \cite{fritz06}. Consequently, the CYGNUS and SKIRTOR models tend to overestimate the AGN and starburst luminosities and underestimate the host galaxy luminosity, when polar dust is not taken into account. A further notable finding is that the stellar mass is significantly underestimated in all models without polar dust. 

\section{ Discussion}\label{sec:discussion}

The presence of polar dust in AGN has been a topic of discussion since the early 90s, with notable contributions by \cite{braatz93} and \cite{cam93}, who detected extended mid-infrared emission to be co-spatial with the ionization cone and the radio jet and outflow in the nearby Seyfert 2 galaxy NGC1068. \cite{asm19} provide a more recent discussion of the ubiquity of polar dust in AGN. \cite{haidar24} recently utilized the high spatial resolution of \textit{JWST} to conduct a detailed study of the polar region of the Seyfert galaxy ESO~428-G14, part of the Galaxy Activity, Torus, and Outflow Survey (GATOS) survey of local AGN. They detected extended mid-infrared emission within 200 pc from the nucleus. This polar structure is co-linear with a radio jet and lies perpendicular to a molecular gas lane that feeds and obscures the nucleus. Its morphology bears a striking resemblance to that of gas ionized by the AGN in the narrow-line region. \cite{haidar24} demonstrate that part of this spatial correspondence is due to contamination within the \textit{JWST} filter bands from strong emission lines. Correcting for the contamination, they determine that the morphology of the dust continuum is more compact, although it is still clearly extended out to $\sim$100 pc. They estimate the temperature of the emitting dust to be $\sim$120K. Through simple models, they also find that the heating of small dust grains by the radiation from the central AGN and/or radiative jet-induced shocks is responsible for the extended mid-infrared emission. Additionally, they conclude that radiation-driven dusty winds from the torus are unlikely to play a significant role.

A number of studies sought to find evidence for the CAT3D-WIND polar wind model of \cite{hk17}, which is in many respects similar to what is proposed in this paper. \cite{isbell21} carried out a survey at $3-5~\mu m$ of 119 local ($z < 0.3$) AGN and found that models including polar winds best reproduced the $3-5~\mu m$ colors, indicating that it is an important component of dusty torus models. \cite{martin21} observed at high angular resolution 13 $z < 0.1$ QSOs in the near-infrared and concluded that the majority are better fitted by a disc plus wind model. \cite{alonso21} compared mid-infrared and submillimetre images of 12 nearby (median 21 Mpc) Seyfert galaxies selected from GATOS with those generated with the CAT3D-WIND model and found good agreement. \cite{garcia22} fitted the SEDs of 24 Seyferts with a range of torus models and found that clumpy disc plus wind models provide the best fits to the nuclear infrared SEDs of Sy1/1.8/1.9 in the near-infrared range.

The fact that polar dust can be satisfactorily modeled with a single-temperature dust component may be an indication that dust is concentrated in an outflow with an $r^{-2}$ density distribution, so the emission is dominated by the inner edge of the outflow. However, \cite{lopez25} concluded that the polar dust in a sample of local AGN (distance $\sim$35 Mpc) they studied with \textit{JWST} at $\sim$75 pc resolution may be distributed in molecular clouds and/or shocked regions across the narrow-line region. The temperature of polar dust in the sample of \cite{lopez25} is estimated to be 132$\pm$7 K, i.e. much lower than the temperature inferred in Arp~299 \citep{matt18} and in the HERUS ULIRGs \citep{smart,varn25}. It may be the case that the origin of polar dust in the sample of \cite{lopez25}, as well as in ESO~428-G14 discussed above, and in the objects we detected polar dust from SED fitting may be different. We note that polar dust with a temperature of the order of 100K is harder to detect with SED fitting, as its emission falls in the same wavelength range as the torus itself, which dominates the emission. In conclusion, the nature of polar dust in W2246$-$0526 and in local Seyfert galaxies studied by \cite{lopez25} may be very different in view of the orders of magnitude difference in infrared luminosity. The polar dust identified in W2246$-$0526 may be more closely related to the component identified in the studies by \cite{alonso21}, \cite{isbell21}, \cite{martin21} and \cite{garcia22}, discussed above.

\cite{vayner25} recently reported the discovery of a dusty outflow in W2246$-$0526 deduced from the identification of several hydrocarbon features detected in absorption using \textit{JWST}/MIRI-MRS observations. The outflow, which is blue-shifted by 5250 km/s from the systemic redshift of the galaxy, is likely accelerated by radiation pressure from the central quasar via trapped infrared photons reprocessed by dust. This discovery is in line with and supports our interpretation that the hot dust component needed to obtain optimal fits to W2246$-$0526's SED may have a polar origin and could be associated to the same dusty outflow seen through hydrocarbon molecules.

W2246$-$0526 has been extensively studied in the literature (e.g. \citealt{assef10,assef15,tsai15,tsai18,tanio18,tanio21}). \cite{tsai15} estimated its bolometric luminosity to exceed $3 \times 10^{14} L_\odot$, while \cite{tanio18} and \cite{tsai18} reported similar values of $3.5 \times 10^{14} L_\odot$ and $3.6\pm0.3 \times 10^{14} L_\odot$, respectively. More recently, \cite{tanio21} reported an infrared ($8-1000~\mu m$) luminosity of $2.2 \times 10^{14} L_\odot$. We note that the luminosities derived from SMART are integrated over $1-1000~\mu m$. Nevertheless, all of the above values are significantly lower than those predicted by our best-fitting models. We attribute this discrepancy to the presence of a heavily obscured AGN that requires a large anisotropy correction, which we have taken into account in our analysis, leading to the high-predicted AGN luminosity. 

Based on kinematic analysis, \cite{tsai18} estimated a black hole mass of $4\pm3.7 \times 10^{9} M_\odot$, about a factor of 3 lower than our estimated range of $1.3-2.3 \times 10^{10} M_\odot$. \cite{tanio18} reported an SFR within the central $\sim$4 arcsec of $560~M_\odot yr^{-1}$, with a factor of 2 uncertainty, and a stellar mass of $2.6 \times 10^{11} M_\odot$, with an uncertainty of a factor of 3. \cite{tanio21} predicted an SFR of $688^{1310}_{-590} M_\odot yr^{-1}$ and a stellar mass of $3^{2.1}_{-1.3} \times 10^{11} M_\odot$. SMART predicts an SFR in the range $358.3-2916~M_\odot yr^{-1}$ and a stellar mass in the range $4.8-5 \times 10^{11} M_\odot$.

Overall, our estimates for the SFR and stellar mass are in agreement with previous estimates within the errors. We note that, as the emission in the rest frame far-infrared is dominated by the AGN torus, the SFR estimate from SMART depends strongly on the far-infrared emissivity of the dust in the torus, which is very uncertain at that redshift. We see, for example, that the model of \cite{sieb15}, which has the highest far-infrared emissivity, predicts the lowest SFR.  

For the black hole estimate, there is more disagreement, but it could be only about a factor of 2, as the error on the kinematical estimate of \cite{tsai18} is very large ($4\pm3.7 \times 10^{9} M_\odot$). One of the most important conclusions of our study is that the black hole mass we infer, assuming the AGN radiates at the Eddington limit, is $\sim$2$-$3 times higher than the more uncertain kinematical estimate of \cite{tsai18}. However, as noted by \cite{tsai18}, the quasar may be super-Eddington with an Eddington ratio of 2.8. Our result could be consistent with the kinematical estimate, if the Eddington ratio is $\sim$2$-$3 times higher. It is notable that \cite{far22}, using the radiative transfer solutions of \cite{efs22}, concluded that 12 of the HERUS ULIRGs are super-Eddington and associated with late-stage mergers. Taken together, these comparisons highlight how our study refines and advances the current understanding of W2246$-$0526.

Taking into account the stellar mass that will be accumulated during the whole current starburst episode, the stellar mass of W2246$-$0526 is predicted to be in the range $4.9-5.1 \times 10^{11} M_\odot$. Assuming the black hole mass is in the range $1.3-2.3 \times 10^{10} M_\odot$ as predicted by the SED analysis, then W2246$-$0526 appears to have a black hole mass slightly above that expected for local elliptical galaxies (\citealt{far23}, see their Fig. 3). If the quasar is super-Eddington with a black hole mass of $\sim 4 \times 10^{9} M_\odot$ as determined by \cite{tsai18}, then W2246$-$0526 lies almost exactly on the $M_{BH}-M_{\text{stellar}}$ relationship of \cite{reines15} of local quiescent elliptical galaxies. These positions relative to local scaling relations do not take into account the possibility of further star formation and/or SMBH accretion episodes, which, given the redshift of W2246$-$0526, are quite possible.

\section{Conclusions}\label{sec:conclusion}

We used multiwavelength photometry, including \textit{JWST} data, to explore the physical properties of the extremely luminous obscured quasar W2246$-$0526 at $z=4.601$, the most luminous galaxy currently known, with the SMART SED fitting code \citep{smart,ascl}, by incorporating four different AGN torus models, in addition to a starburst and a spheroidal host galaxy. Apart from the combination with the AGN torus model of \cite{sieb15}, we also fitted the data with the same combinations of models but with the addition of a hot dust component that we attribute to polar dust emission in the context of a torus geometry. All models statistically prefer the presence of this component, suggesting it is required in order to account for the rest-frame mid-infrared spectrum of W2246$-$0526. Our main conclusions are summarized as follows:

\begin{enumerate}

\item We find that a combination of an AGN torus, a starburst and a spheroidal component fails to fit the mid-infrared spectrum. Taking into account extinction by the host galaxy does not help to resolve this problem.

\item 
Absorption and re-radiation by dust in the polar regions of the AGN torus, in addition to an edge-on torus, is proposed as the most plausible scenario for explaining the observed infrared SED of W2246$-$0526.

\item
We find that the best fit is provided by the smooth CYGNUS tapered disc model, followed closely by the SKIRTOR two-phase model, both with polar dust enabled. The smooth flared disc model of \cite{fritz06} with polar dust enabled provides a satisfactory solution, albeit with a slightly higher minimum reduced $\chi^2$.

\item
Our method predicts an AGN fraction in the range $72-81$ per cent, an SFR in the range $358.3-2916~M_\odot yr^{-1}$ and a stellar mass of $4.8-5 \times 10^{11} M_\odot$. The predicted polar dust luminosity is between $1.6-1.7 \times 10^{14} L_\odot$ and the AGN luminosity is in the range $4.2-7.2 \times 10^{14} L_\odot$. The covering factor of polar dust is predicted to be in the range $24-38$ per cent. The estimated black hole mass of W2246$-$0526 is between $1.3-2.3 \times 10^{10} M_\odot$.

\item
Our estimates of the SFR and stellar mass are consistent with earlier estimates within the uncertainties. However, our estimate of the AGN luminosity, and therefore the black hole mass, is higher by a factor of $\sim$2$-$3 compared with previous estimates. W2246$-$0526 may therefore  have an Eddington ratio $\sim$2$-$3 times higher than the value of 2.8 estimated by \cite{tsai18}.

\item 
Our analysis shows that SED fitting may be a very powerful technique for detecting potential polar dust in high-redshift AGN, which in turn may be associated to the presence of fast, dusty outflows in obscured sources \citep{vayner25} not directly detected by mid-infrared imaging and spectroscopy.

\end{enumerate}

\section*{Acknowledgements}

We thank an anonymous referee for useful comments and suggestions. CV, AE and TDS acknowledge support from the projects CYGNUS (contract number 4000126896) and CYGNUS+ (contract number 4000139319) funded by the European Space Agency.

\section*{Data Availability}

The data underlying this article are available in the article or are publicly available in the literature.

%%%%%%%%%%%%%%%%%%%%%%%%%%%%%%%%%%%%%%%%%%%%%%%%%%

%%%%%%%%%%%%%%%%% APPENDICES %%%%%%%%%%%%%%%%%%%%%

%%%%%%%%%%%%%%%%%%%%%%%%%%%%%%%%%%%%%%%%%%%%%%%%%%

\appendix

\section{Quality tests for SED fitting}

In this Appendix, we demonstrate how we assess the quality of the SED fitting, by implementing a comprehensive suite of tests. These include evaluating the goodness of fit, using statistical indicators such as the minimum reduced $\chi^2$, and performing a $\chi^2$ difference test for nested models to compare the quality of fits between nested models. To take into account the number of data points and penalize model complexity, we also employ the BIC as a complementary metric. Through this rigorous evaluation, we obtain more reliable constraints on the physical properties of W2246$-$0526.

\subsection{Minimum reduced $\chi^2$}
There is significant uncertainty in the models, which is primarily due to uncertainties regarding the dust properties, the geometry, numerical errors etc. The uncertainty in the observed data is driven primarily by the calibration uncertainty of the instruments. We therefore define the $\chi^{2}$ statistic we use to compare the observed and the model SEDs as:
\begin{equation}\label{chi^2}
\chi^{2}=\sum_{n}\left[\frac{\left(y_{n}-\mu_{n}\right)^{2}}{(\sigma_y~y_{n})^{2}+(\sigma_{\mu}~\mu_{n})^{2}}\right]~~,
\end{equation}
where $y_n$ and $\mu_n$ are the observed data and the corresponding model value respectively, $\sigma_y$ is the uncertainty of the data and $\sigma_{\mu}$ is the uncertainty of the model.

For the $\chi^{2}$ calculation, we adopted a 15 per cent uncertainty for the model SEDs and a 20 per cent uncertainty for the observed data. The minimum reduced $\chi^{2}$, $\chi^{2}_{min, \nu}$, can then be obtained by:
\begin{equation}
{\chi^{2}_{min,\nu}=\frac{\chi^{2}_{min}}{\nu}}~~,
\end{equation}
where the degrees of freedom, $\nu=n-m$, equal the number of data points $n$ minus the number of fitted parameters $m$. In Table \ref{tab:fitted} we list the minimum reduced $\chi^{2}$ values provided by the fits with the four different torus models, with and without the addition of the polar dust component.

\subsection{$\chi^2$ difference test for nested models}\label{chi_2_diff}
The $\chi^2$ difference test is a statistical test to assess whether the addition of model parameters leads to a statistically significant improvement in the fit to the data. This test helps determine whether added parameters improve model fit sufficiently to justify increased complexity. This ensures that the preferred model provides a robust representation of the physical properties of W2246$-$0526, without overfitting the observational data. To compare two nested models using the $\chi^2$ difference, the test statistic is given by:
\begin{equation}
\Delta \chi^2=\chi^2_{\text{simple}}-\chi^2_{\text{complex}}~~,
\end{equation}
which follows a $\chi^2$ distribution with
\begin{equation}
\Delta\nu=m_{\text{complex}}-m_{\text{simple}}
\end{equation}
degrees of freedom under the null hypothesis. A low $\text{p-value}$ ($\text{p}<10^{-4}$) indicates that the additional parameters in the complex model lead to a statistically significant improvement in the fit. In Table \ref{tab:test} we list the results of the $\chi^2$ difference test for each torus model, with and without the polar dust component.

\subsection{BIC for model comparison}\label{BIC}
The BIC is also used to compare competing models, by balancing the quality of the fit against model complexity. This criterion is particularly useful, as it penalizes the inclusion of unnecessary parameters, helping to avoid overfitting. Unlike the $\chi^2$ difference test, the BIC also takes into account the number of independent data points available. The BIC for a given model is defined as:
\begin{equation}
\mathrm{BIC}=m\ln(n)-2\ln(\hat{L})~~,
\end{equation}
where $m$ is the number of free model parameters, $n$ is the number of independent data points and $\hat{L}$ is the maximized value of the likelihood function, i.e.
\begin{equation}
\hat{L}=P(D \mid\ \hat{\Theta},M)~~,
\end{equation}
where $D$ is the observed data, $\{\hat{\Theta}\}$ are the parameter values that maximize the likelihood function and $M$ is the model. Since our likelihood function is a Gaussian, the log-likelihood is related to the $\chi^2$ statistic by:
\begin{equation}
\ln(\hat{L})=-\frac{1}{2}\chi^2+\text{C}~~,
\end{equation}
where $\text{C}$ is a constant. We can therefore approximate BIC for model comparison as:
\begin{equation}
\mathrm{BIC}\approx\chi^2+m\ln(n)~~.
\end{equation}
A lower BIC value indicates a preferred model. To compare two models, the difference
\begin{equation}
\Delta\mathrm{BIC}=\mathrm{BIC}_{\text{complex}}-\mathrm{BIC}_{\text{simple}}
\end{equation}
is evaluated, with $\Delta\mathrm{BIC}>0$ favoring the simpler model and $\Delta\mathrm{BIC}<0$ favoring the more complex model. In Table \ref{tab:test} we list the results of the BIC for each torus model, with and without the polar dust component. 

For the interpretation of the strength of evidence, we follow the evidence scale of \cite{kass95}, interpreting differences in BIC as approximations to twice the logarithm of the Bayes factor:

\begin{tabular}{ll}
\hline
$|\Delta \mathrm{BIC}|$ & Interpretation \\
\hline
0$-$2 & Weak evidence \\
2$-$6 & Positive evidence \\
6$-$10 & Strong evidence \\
$>10$ & Very strong evidence \\
\hline
\end{tabular}

%%%%%%%%%%%%%%%%%%%%%%%%%%%%%%%%%%%%%%%%%%%%%%%%%%

% Don't change these lines
\bsp	% typesetting comment
\label{lastpage}

\begin{thebibliography}{}

\bibitem[Alonso-Herrero et al.(2021)]{alonso21} 
Alonso-Herrero A., Garc{\'\i}a-Burillo S., H{\"o}nig S.~F. et al, 2021, \aap, 652, A99

\bibitem[Asmus(2019)]{asm19} 
Asmus D., 2019, \mnras, 489, 2177

\bibitem[Assef et al.(2010)]{assef10} 
Assef R.~J., Kochanek C.~S., Brodwin M. et al, 2010, \apj, 713, 970 

\bibitem[Assef et al.(2015)]{assef15} 
Assef R.~J., Eisenhardt P.~R.~M., Stern D. et al, 2015, \apj, 804, 27 

\bibitem[Boquien et al.(2019)]{boq19} 
Boquien M., Burgarella D., Roehlly Y. et al., 2019, \aap, 622, A103

\bibitem[Braatz et al.(1993)]{braatz93} 
Braatz J.~A., Wilson A.~S., Gezari D.~Y. et al., 1993, \apjl, 409, L5

\bibitem[Bridge et al.(2013)]{bridge13} 
Bridge C.~R., Blain A., Borys C.~J.~K. et al., 2013, \apj, 769, 91 

\bibitem[Bruzual \& Charlot(1993)]{bru93} 
Bruzual  G. \& Charlot S., 1993, \apj, 405, 538 

\bibitem[Bruzual \& Charlot(2003)]{bru03} 
Bruzual G. \& Charlot S., 2003, \mnras, 344, 1000 

\bibitem[Burtscher et al.(2013)]{burt13} 
Burtscher L., Meisenheimer K., Tristram K.~R.~W. et al., 2013, \aap, 558, A149

\bibitem[Cameron et al.(1993)]{cam93} 
Cameron M., Storey J.~W.~V., Rotaciuc V. et al., 1993, \apj, 419, 136

\bibitem[da Cunha et al.(2008)]{dac08} 
da Cunha E., Charlot S. \& Elbaz D., 2008, \mnras, 388, 1595

\bibitem[D{\'\i}az-Santos et al.(2016)]{tanio16} 
D{\'\i}az-Santos T., Assef R.~ J., Blain A.~W. et al., 2016, \apjl, 816, L6

\bibitem[D{\'\i}az-Santos et al.(2018)]{tanio18} 
D{\'\i}az-Santos T., Assef R.~ J., Blain A.~W. et al., 2018, Science, 362, 1034

\bibitem[D{\'\i}az-Santos et al.(2021)]{tanio21} 
D{\'\i}az-Santos T., Assef R.~ J., Eisenhardt P.~R.~M. et al., 2021, \aap, 654, A37

\bibitem[D{\'\i}az-Santos et al.(2025)]{tanio25} 
D{\'\i}az-Santos T., Lai T.~S.~Y., Finnerty L. et al., 2025, Astrophysics Source Code Library, record ascl:2501.001

\bibitem[Efstathiou \& Rowan-Robinson(1995)]{efstathiou95} 
Efstathiou A. \&  Rowan-Robinson M., 1995, \mnras, 273, 649

\bibitem[Efstathiou, Hough \& Young(1995)]{efs95} 
Efstathiou A., Hough J.~H., Young S., 1995, \mnras, 277, 1134

\bibitem[Efstathiou et al.(2000)]{efstathiou00} 
Efstathiou A., Rowan-Robinson M. \& Siebenmorgen R., 2000, \mnras, 313, 734

\bibitem[Efstathiou \& Rowan-Robinson(2003)]{efs03} 
Efstathiou A. \& Rowan-Robinson M., 2003, \mnras, 343, 322

\bibitem[Efstathiou(2006)]{efstathiou06} 
Efstathiou A., 2006, \mnras, 371, L70

\bibitem[Efstathiou \& Siebenmorgen(2009)]{efstathiou09} 
Efstathiou A. \& Siebenmorgen R., 2009, \aap, 502, 541

\bibitem[Efstathiou et al.(2013)]{efstathiou13} 
Efstathiou A., Christopher N., Verma A. \& Siebenmorgen R., 2013, \mnras, 436, 1873

\bibitem[Efstathiou et al.(2014)]{efs14} 
Efstathiou A., Pearson C., Farrah D. et al., 2014, \mnras, 437, L16

\bibitem[Efstathiou et al.(2021)]{efs21} 
Efstathiou A., {Ma{\l}ek} K., {Burgarella} D. et al., 2021, \mnras, 503, L11

\bibitem[Efstathiou et al.(2022)]{efs22} 
Efstathiou A., Farrah D., Afonso J. et al., 2022, \mnras, 512, 5183

\bibitem[Eisenhardt et al.(2012)]{eins12} 
Eisenhardt P.~R.~M., Wu J., Tsai C.-W. et al., 2012, \apj, 755, 173

\bibitem[Fan et al.(2020)]{fan20} 
Fan L., Chen W., An T. et al., 2020, \apjl, 905, L32

\bibitem[Farrah et al.(2013)]{far13} 
Farrah D., Lebouteiller V., Spoon H.~W.~W. et al., 2013, \apj, 776, 38

\bibitem[Farrah et al.(2017)]{far17} 
Farrah D., Petty S., Connolly B. et al., 2017, \apj, 844,106

\bibitem[Farrah et al.(2022)]{far22} 
Farrah D., Efstathiou A., Afonso J. et al., 2022, \mnras, 513, 4770

\bibitem[Farrah et al.(2023)]{far23} 
Farrah D., Petty S., Croker K.~S. et al., 2023, \apj, 943, 133

\bibitem[Farrah et al.(2026)]{far26} 
Farrah D., Ejercito K., Efstathiou A. et al., 2026, \apj, 997, 150

\bibitem[Fritz et al.(2006)]{fritz06} 
Fritz J., Franceschini A. \& Hatziminaoglou E., 2006, \mnras, 366, 767

\bibitem[G{\'a}mez Rosas et al.(2022)]{gamez22}
G{\'a}mez Rosas V., Isbell J.~W., Jaffe W. et al. 2022, \nat, 602, 403 

\bibitem[Garc{\'\i}a-Bernete et al.(2022)]{garcia22} 
Garc{\'\i}a-Bernete I., Gonz{\'a}lez-Mart{\'\i}n O., Ramos Almeida C. et al. 2022, \aap, 667, A140

\bibitem[Haidar et al.(2024)]{haidar24} 
Haidar H., Rosario D.~J., Alonso-Herrero A. et al., 2024, \mnras, 532, 4645

\bibitem[H{\"o}nig et al.(2012)]{honig12} 
H{\"o}nig S.~F., Kishimoto M., Antonucci R. et al., 2012, \apj, 755, 149

\bibitem[H{\"o}nig et al.(2013)]{honig13} 
H{\"o}nig S.~F., Kishimoto M., Tristram K.~R.~W. et al., 2013, \apj, 771, 87

\bibitem[H{\"o}nig \& Kishimoto(2017)]{hk17} 
H{\"o}nig S.~F. \& Kishimoto M., 2017, \apjl, 838, L20

\bibitem[H{\"o}nig(2019)]{honig19} 
H{\"o}nig S.~F., 2019, \apj, 884, 171

\bibitem[Isbell et al.(2021)]{isbell21}
Isbell J.~W., Burtscher L., Asmus D. et al., 2021, \apj, 910, 104

\bibitem[Isbell et al.(2022)]{isbell22}
Isbell J.~W., Meisenheimer K., Pott J.~U. et al., 2022, \aap, 663, A35

\bibitem[Isbell et al.(2023)]{isbell23} 
Isbell J.~W., Pott J.~U., Meisenheimer K. et al., 2023, \aap, 678, A136

\bibitem[Jaffe et al.(2004)]{jaffe04} 
Jaffe W., Meisenheimer K., R{\"o}ttgering H.~J.~A. et al., 2004, \nat, 429, 47

\bibitem[Kass \& Raftery(1995)]{kass95} 
Kass R.~E. \& Raftery A.~E., 1995, J. Am. Stat. Assoc., 90, 773

\bibitem[Leftley et al.(2018)]{leftley18} 
Leftley J., Tristram K., H{\"o}nig S. et al. 2018, \apj, 862, 17 

\bibitem[L{\'o}pez-Gonzaga et al.(2014)]{lopez14} 
L{\'o}pez-Gonzaga N., Jaffe W., Burtscher L. et al., 2014, \aap, 565, A71

\bibitem[L{\'o}pez-Gonzaga \& Jaffe(2016)]{lopez16a} 
L{\'o}pez-Gonzaga N. \& Jaffe W., 2016, \aap, 591, A128

\bibitem[L{\'o}pez-Gonzaga et al.(2016)]{lopez16b} 
L{\'o}pez-Gonzaga N., Burtscher L., Tristram K.~R.~W et al., 2016, \aap,  591, A47

\bibitem[Lopez-Rodriguez et al.(2025)]{lopez25} 
Lopez-Rodriguez E., Ramos Almeida C., Pereira-Santaella M. et al., 2025, \apj, 994, 206

\bibitem[Lyu \& Rieke (2022)]{lyu22} 
Lyu J. \& Rieke G.~H., 2022\ \apjl, 940, L31

\bibitem[Mart{\'\i}nez-Paredes et al.(2021)]{martin21} 
Mart{\'\i}nez-Paredes M., Gonz{\'a}lez-Mart{\'\i}n O., HyeongHan K. et al., 2021, \apj, 922, 157

\bibitem[Mattila et al.(2018)]{matt18} 
Mattila S., P{\'e}rez-Torres M. {\'A}., Efstathiou A. et al., 2018, Science, 361, 482

\bibitem[Noll et al.(2009)]{noll09} 
Noll S., Burgarella D., Giovannoli E. et al., 2009, \aap, 507, 1793

\bibitem[Raban et al.(2009)]{raban09}
Raban D., Jaffe W., R{\"o}ttgering H. et al., 2009, \mnras, 394, 1325

\bibitem[Reines \& Volonteri(2015)]{reines15}
Reines A.~E. \& Volonteri M., 2015, \apj, 813, 82

\bibitem[Reynolds et al.(2022)]{reyn22} 
Reynolds T.~M., Mattila S., Efstathiou A. et al., 2022, \aap, 664, A158

\bibitem[Siebenmorgen et al.(2015)]{sieb15} 
Siebenmorgen R., Heymann F. \& Efstathiou A., 2015, \aap, 583, A120 

\bibitem[Stalevski et al.(2012)]{sta12} 
Stalevski M., Fritz J., Baes M. et al., 2012, \mnras, 420, 2756 

\bibitem[Stalevski et al.(2016)]{stal16} 
Stalevski M., Ricci C., Ueda Y. et al., 2016, \mnras, 458, 2288

\bibitem[Stalevski et al.(2017)]{stal17} 
Stalevski M., Asmus D. \& Tristram K.~R.~W., 2017, \mnras, 472, 3854 

\bibitem[Stalevski et al.(2019)]{stal19} 
Stalevski M., Tristram K.~R.~W. \& Asmus D., 2019, \mnras, 484, 3334

\bibitem[Tristram et al.(2009)]{tristram09}
Tristram K.~R.~W., Raban D., Meisenheimer K. et al., 2009, \aap, 502, 67

\bibitem[Tristram et al.(2014)]{tristram14}
Tristram K.~R.~W., Burtscher L., Jaffe W. et al., 2014, \aap, 563, A82

\bibitem[Tsai et al.(2015)]{tsai15}
Tsai C.-W., Eisenhardt P.~R.~M., Wu J. et al., 2015, \apj, 805, 90

\bibitem[Tsai et al.(2018)]{tsai18}
Tsai C.-W., Eisenhardt P.~R.~M, Jun H.~D. et al., 2018, \apj, 868, 15

\bibitem[Varnava \& Efstathiou(2024a)]{smart} 
Varnava C. \& Efstathiou A., 2024a, \mnras, 531, 2304

\bibitem[Varnava \& Efstathiou(2024b)]{ascl} 
Varnava C. \& Efstathiou A., 2024b, Astrophysics Source Code Library, record ascl:2406.003

\bibitem[Varnava et al.(2024)]{varn24} 
Varnava C., Efstathiou A. \& Farrah D., 2024, \mnras, 534, 2585

\bibitem[Varnava et al.(2025)]{varn25} 
Varnava C., Efstathiou A., Farrah D. \& Rigopoulou D., 2025 \mnras, 538, 426

\bibitem[Vayner et al.(2025)]{vayner25} 
Vayner A., D{\'\i}az-Santos T., Ferkinhoff C.~D. et al., 2025, preprint (arXiv:2510.09870) 

\bibitem[Wright et al.(2010)]{wright10} 
Wright E.~L., Eisenhardt P.~R.~M., Mainzer A.~K. et al., 2010, \aj, 140, 1868

\bibitem[Wu et al.(2012)]{wu12}
Wu J., Tsai C.-W., Sayers J. et al., 2012, \apj, 756, 96

\bibitem[Young et al.(1996)]{young96} 
Young S., Hough J.~H., Efstathiou A. et al., 1996, \mnras, 281, 1206

\end{thebibliography}
\end{document}